# Altitude-Dependent Near-Source Spectral Filtering of Meteor Infrasound Above 80 km and Consequences for Period-Based Energy Estimates

Elizabeth A. Silber[1,*], Igor P. Chunchuzov[2], Oleg E. Popov[2], Sergey N. Kulichkov[2,3]

[1]Sandia National Laboratories, 1515 Eubank Blvd. NE, Albuquerque, NM, 87123, US

[2]Obukhov Institute of Atmospheric Physics, Russian Academy of Sciences, Moscow, 119017 Russia

[3]Moscow State University, Moscow, 119991 Russia

**Accepted for publication in Pure and Applied Geophysics (27-Jul-2026)**

*Corresponding author: esilbe@sandia.gov

**Abstract**

Infrasound signal period is widely used as a proxy for source energy in bolide studies, but the reliability of period–yield relations for small, high-altitude regional meteors has not been systematically evaluated. We analyze 90 infrasound detections from well-constrained regional meteoroid events (source altitudes 20–111 km, ranges 47–268 km) and demonstrate a strong, monotonic decrease in receiver dominant frequency with increasing source altitude (Spearman $r_s$ = −0.629, $p = 3.3 \times 10^{-11}$). No detections above 80 km retain dominant frequencies exceeding 3 Hz, and 75% of detections above 100 km are dominated by sub-1 Hz content. Partial correlation analysis indicates this is primarily an altitude effect, not a propagation distance artifact. Near-source dissipation modeling using the generalized Burgers equation supports a physical mechanism: the exponential increase in kinematic viscosity with altitude drives the acoustic Reynolds number downward, imposing frequency-dependent molecular absorption that selectively attenuates high-frequency content within the first 5 km below the source. Our results suggest that this atmospheric low-pass filter systematically modulates the observed period and can bias period-based energy estimates upward by one to two orders of magnitude for sources above 80–90 km when uncorrected period–yield relations are applied. These findings are relevant to any high-altitude infrasound source, including space debris and controlled reentries.

## 1. Introduction

Meteoroids entering Earth's atmosphere generate a broad suite of observable signatures, including optical emission, ionization, and acoustic waves (Bronshten, 1983; Ceplecha et al., 1998; Silber et al., 2018). The combined interpretation of these signatures is fundamental to small body science, atmospheric entry physics, and impact hazard assessment (Brown et al., 2002; Devillepoix et al., 2020; Jenniskens et al., 2000; Popova et al., 2013; Silber et al., 2009). Among these modalities, infrasound, which consists of low frequency acoustic waves below 20 Hz (Evans et al., 1972), offers a uniquely scalable capability. Because atmospheric attenuation for these long wavelengths is minimal, infrasound provides passive detection of energetic atmospheric sources across regional to global distances (Campus and Christie, 2009). This enables event characterization even when optical or space-based measurements are absent or incomplete (Scamfer et al., 2026; Silber and Brown, 2019; Wilson et al., 2025). At the same time, bolide infrasound remains a demanding inference problem because the source is elevated, moving, and often fragmenting, while the recorded signal reflects not only source processes but also the cumulative effects of propagation through a structured and time varying atmosphere. Recent syntheses have indicated that improving yield and source characterization hinges on resolving persistent interpretive inconsistencies, particularly variability in measured signal periods, and on leveraging larger, systematically analyzed event samples to separate intrinsic source physics from extrinsic propagation and station effects (Silber et al., 2025c).

A major reason infrasound signal period has received sustained attention is its historical role as a practical proxy for energetics (ReVelle, 1997). Period–yield (period–energy) relations, adapted from nuclear test monitoring and subsequently recalibrated for bolides, remain widely used because they can be applied rapidly and do not require dense station coverage or detailed source modeling (Ens et al., 2012; Gi and Brown, 2017; ReVelle, 1997). However, recent work has demonstrated that a single global period–energy relation can conceal substantial variability introduced by event geometry, atmospheric conditions, and the altitude dependent nature of bolide energy deposition and fragmentation behavior. Partitioning by factors such as altitude, entry angle, distance, and fragmentation type can shift inferred slopes and intercepts. This shows that period-based inference is not governed by a single universal mapping from period to yield, but rather by a coupled source–atmosphere system whose controlling parameters can change from event to event (Silber et al., 2025c). These findings therefore motivate deeper scrutiny of when and why period-based estimates are reliable.

Propagation through the real atmosphere is a particularly important and sometimes under-acknowledged driver of period variability. Fine-scale wind and temperature structure can modulate amplitude and travel time, contributing directly to errors in source characterization if not accounted for (Averbuch et al., 2022; Chunchuzov and Kulichkov, 2020; Chunchuzov et al., 2011; Chunchuzov et al., 2025; Ostashev and Wilson, 1997; Wilson, 1996). Moreover, recent modeling and observational comparisons have noted that waveform shape, duration, and characteristic period can change systematically with source altitude and range in the presence of fine scale layered structure (Chunchuzov et al., 2025; Chunchuzov et al., 2026). Critically, the dominant frequency measured at the receiver reflects not only the emitted spectrum but also the frequency dependent transmission properties of the atmosphere. This point is especially consequential because source energy is often inferred, explicitly or implicitly, from received spectral content and dominant period (ReVelle, 1997). In settings where atmospheric structure modulates the observed oscillation period, straightforward interpretation can bias energetic inferences upward, even when the underlying source is unchanged (e.g., Silber et al., 2026a).

These issues become acute for small, regional meteoroids whose infrasound sources frequently occur at very high altitudes (Moreno-Ibáñez et al., 2018; Silber and Brown, 2014). Classical expectations for shock formation can break down in the presence of intense ablation (Silber et al., 2018), and empirical evidence indicates that meteoroids can generate shocks at altitudes significantly higher than anticipated under simplified continuum flow assumptions, including occasional detections originating above 100 km (Brown et al., 2007; Moreno-Ibáñez et al., 2018; Silber et al., 2026b). High altitude sources are therefore not exceptional outliers but a recurring part of the regional meteor infrasound problem. Yet this regime is exactly where thermodynamic state, characterized by low density and low pressure, and atmospheric structure are expected to preferentially suppress higher frequency content during downward propagation, yielding ground recordings that are increasingly dominated by longer periods (Silber and Brown, 2014). The resulting signals can resemble those expected from substantially larger, lower altitude sources, complicating direct physical interpretation and motivating a cautious stance on using period as a yield proxy without additional constraints (e.g., Silber et al., 2026b).

Here we leverage a well-constrained, curated regional optical–infrasound dataset (Silber et al., 2025a; Silber et al., 2025b) to motivate a physics-rooted reassessment of how dominant period behaves for very high-altitude regional meteor sources. The overarching objective of this paper is to clarify, on a physics-informed basis, how and why very high-altitude regional meteor sources can produce systematically lower frequency ground signals, and to delineate the resulting limitations on period-based yield inference and related interpretations for this abundant class of events. We treat this as an exploratory, first-order study focused on the altitude dependence; other factors (e.g., station noise, background noise, and signal/source strength) can influence detectability and are not explicitly treated here. The paper is organized as follows. Section 2 reviews the physical basis of period–yield scaling and expected limitations for high altitude sources, Section 3 describes the dataset and analysis methods, Section 4 presents the results and discussion, and Section 5 provides our conclusions.

## 2. Theory and Background

### 2.1 Period–Energy Relations as Interpretive Context

A recurring goal in bolide infrasound research is to relate what is measured at the ground, typically a waveform and its spectrum, to physically meaningful source properties such as energy deposition, fragmentation behavior, and source altitude (e.g., McFadden et al., 2021; ReVelle, 1976; Silber and Brown, 2014). In operational and research practice, the dominant period (or equivalently the dominant frequency) is commonly used for this purpose because it is readily measured from infrasonic records and has long been used to characterize atmospheric explosions (ReVelle, 1997).

Period–yield (or period–energy) relations link a characteristic period to an energetic scale through an empirically calibrated power law:

$$\log_{10}E = A\log_{10}T_{dsp} + B, \quad (1)$$

where $E$ is the total energy in kt of TNT (1 kt TNT (trinitrotoluene) = $4.184\cdot10^{12}$ J), $T_{dsp}$ is the dominant signal period in seconds, and the coefficients A and B are empirically derived from events where energetics is independently constrained. Several standard relations for bolides have been established, including those by ReVelle (1997), Ens et al. (2012), Gi and Brown (2017), and Silber et al. (2025c). The general form is presented here for interpretive context, to illustrate why period has been treated as a proxy for energetic scale, and therefore why any systematic modulation of the receiver dominant period has direct consequences for meteoroid energetics inferred from infrasound.

The use of infrasonic signal period as a proxy for source strength originates from the similarity properties of blast waves (e.g., Sachdev, 1972; Sakurai, 1965). In an idealized setting, a point-like explosive release in a homogeneous atmosphere, the early-time flow is strongly nonlinear near the source, but the disturbance rapidly decays with range and transitions into a weakly nonlinear acoustic shock, often idealized as an N-wave (Dumond et al., 1946; Landau, 1945). The emitted waveform is associated with a characteristic length scale that depends on the energy release and the ambient state, and that length scale maps to a characteristic timescale (hence, a characteristic period) through division by the local sound speed.

In practice, meteoroid entry complicates the ideal picture because energy deposition is distributed along a path and may be punctuated by fragmentation; accordingly, period-based inference is best viewed as relating the observed period to an effective acoustic source region sampled by a given receiver (Silber et al., 2025c; Silber et al., 2026b). This is one reason why multiple published period–energy relations exist in the literature with differing slopes (A ≈ 3.3–3.75) and intercepts. For reference, **Figure 1** compares representative published period–energy relations used in bolide infrasound studies and serves as a reminder that "period-to-energy" mappings depend on calibration choices and event populations.

The steep power-law exponents are a critical feature: because energy scales as period raised to approximately the 3.3rd–3.75th power, even modest changes in the observed period map to large changes in inferred energy. A factor of 2 in period corresponds to a factor of ~10–13 in energy, and a factor of 3 corresponds to ~37–62, depending on the specific relation used. This nonlinear amplification is what makes any systematic atmospheric modulation of the receiver period consequential for energetic inference.

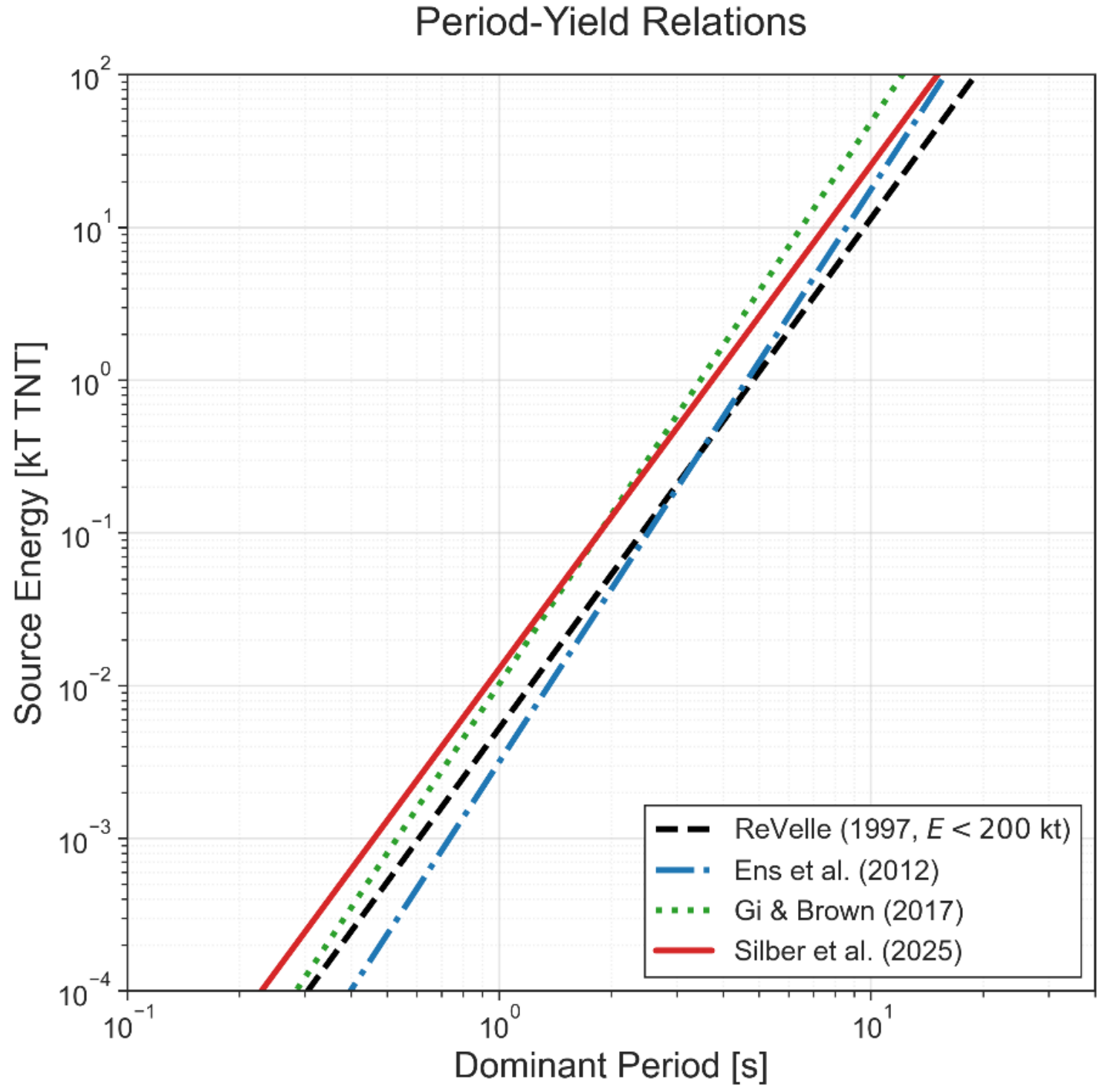

**Figure 1:** Comparison of commonly used empirical period–energy relations for bolide infrasound. The curves represent the relations published by ReVelle (1997), Ens et al. (2012), Gi and Brown (2017), and Silber et al. (2025c).

### 2.2. Source Energy, Scale, and the Role of Observational Distance

The interpretation of infrasonic spectra is governed by both the intrinsic energy of the source and the observational scale. Large, energetic bolide events generate massive disturbances with substantial low-frequency content that is physically capable of traveling thousands of kilometers (e.g., Arrowsmith et al., 2021; Ens et al., 2012; ReVelle, 1997; Silber et al., 2011). For these major events detected at global ranges, frequency-dependent attenuation preferentially strips away higher-frequency spectral components, while sub-hertz and low-hertz waves propagate with minimal dissipative losses. In this far-field regime, the recorded signal is intrinsically low-frequency dominated, and this stable portion of the spectrum serves as the foundational calibration point for established period-yield relations. For large bolides detected at global ranges via stratospheric waveguide propagation, the period measured at the receiver is largely insensitive to the source altitude because the low-frequency content that dominates the far-field signal was never significantly affected by the near-source environment. It should be noted that thermospheric propagation paths, where signals traverse the upper atmosphere, can introduce additional attenuation and spectral modulation (Sutherland and Bass, 2004); however, the majority of bolide detections used to calibrate period-yield relations correspond to stratospheric returns (Ens et al., 2012; Silber et al., 2025c).

Regional observations, involving centimeter-scale meteoroids, differ in a fundamental physical sense. These small sources generate signals naturally richer in high-frequency content, with characteristic periods $T_{dsp}$ of order 0.1–1 s (corresponding to dominant frequencies of 1–10 Hz) rather than the multi-second periods of large bolides (Silber and Brown, 2014). At ranges of tens to a few hundred kilometers, the propagation path is short enough that these higher frequencies can remain observable at the receiver, provided they survive both the near-source environment and the downward path through the atmosphere. Regional datasets thus occupy a diagnostically vital middle ground: they are close enough to the source for the spectral content to retain its informative high-frequency structure, yet far enough that the waveforms reflect realistic atmospheric modulation rather than purely near-field hydrodynamics.

The high-frequency content of regional meteor signals is simultaneously what makes them scientifically valuable and what makes them vulnerable to atmospheric filtering. The vulnerability arises because the atmospheric properties that control dissipation, particularly molecular viscosity, change by orders of magnitude between the lower atmosphere and the upper mesosphere where many of these sources originate (Sutherland and Bass, 2004).

The contrast between large bolides and small meteoroids is reinforced by their characteristic energy deposition altitudes. Large bolides, owing to their substantial mass and momentum, penetrate deep into the atmosphere before depositing the bulk of their kinetic energy, typically in the 20-50 km altitude range (e.g., Brown et al., 2002; Popova et al., 2013). While large bolides generate shock waves along their entire trajectory, the dominant infrasonic signal, which carries the most energy and is detected at global ranges, typically originates from the lower-altitude region of peak energy deposition where molecular viscosity is negligible. Small, centimeter- to decimeter-scale meteoroids, by contrast, span a much broader range of energy deposition altitudes. Their smaller sizes mean they decelerate more rapidly and can ablate and fragment anywhere from below 60 km to well above 100 km, depending on entry velocity, composition, and structure (Ceplecha et al., 1998; Moreno-Ibáñez et al., 2017; Moreno-Ibáñez et al., 2018; Silber et al., 2026b). As a result, small meteoroids routinely generate acoustic sources in the altitude regime where near-source dissipation is most severe, a regime that large bolides rarely sample.

**2.3 Atmospheric Viscosity and the Near-Source Dissipation Problem**

The primary physical property connecting source altitude to spectral content is the kinematic molecular viscosity $\nu$, which increases approximately exponentially with altitude due to the decrease in atmospheric density. At sea level, $\nu \approx 1.5 \times 10^{-5}$ m$^2$/s. At 80 km, $\nu \approx 1.3$ m$^2$/s, an increase of approximately $10^5$. At 100 km, $\nu \approx 130$ m$^2$/s, an increase of approximately $10^7$ relative to the ground (Gossard and Hooke, 1975; Sutherland and Bass, 2004). This rapid increase has profound consequences for the propagation of finite-amplitude acoustic waves in the upper atmosphere.

The competition between nonlinear wave steepening and viscous dissipation in a propagating N-wave is quantified by the acoustic Reynolds number (Rudenko and Soluyan, 1977):

$$Re = \frac{v_0 c}{\nu \omega_0}, \tag{2}$$

where $v_0$ is the radial velocity amplitude at a reference distance from the source, $c$ is the local sound speed, and $\omega_0 = \frac{2\pi}{T_0} = 2\pi f_0$ is the angular frequency of the main spectral maximum of the N-wave near the source, with $T_0$ being the initial signal duration and $f_0$ = $1/T_0$ the corresponding dominant frequency. For a given source strength (fixed $v_0$) and frequency, $Re$ is inversely proportional to $\nu$ and therefore decreases exponentially with altitude. Physically, $Re$ measures the relative importance of finite-amplitude nonlinear steepening and molecular absorption during the early evolution of the N-wave. Equivalently, it can be interpreted as proportional to the ratio between the molecular dissipation length scale, $\alpha_m^{-1}$, and the nonlinear shock-formation distance, $x_{\text{non}}$, up to an order-unity coefficient involving the nonlinear parameter. Thus, large $Re$ indicates that nonlinear steepening acts over a shorter distance than molecular absorption, while small $Re$ indicates that molecular absorption controls the spectral evolution before appreciable nonlinear steepening develops. This single parameter defines two fundamentally different propagation regimes, $Re$ >> 1 and $Re$ << 1. Here and throughout the paper, ‘nonlinear effects’ refers to finite-amplitude wave processes, including shock-front steepening, N-wave stretching, and harmonic generation, whereas molecular/thermoviscous attenuation is treated as the competing dissipative process.

When $Re$ >> 1, finite-amplitude nonlinear effects dominate over molecular absorption in the near-source region. The N-wave can undergo nonlinear stretching and a modest shift of spectral energy toward lower frequencies, while shock-front steepening and harmonic generation help maintain a sharp waveform. For the small acoustic Mach numbers considered here, this near-source modification is limited over the first few kilometers of propagation. As the wave descends into denser air and weakens with geometric spreading, subsequent spectral evolution becomes much smaller, and the signal propagates primarily through refraction, geometric spreading, and scattering. In this regime, the measured period retains a closer correspondence to the source scale and geometry than in the $Re \ll 1$ molecular-dissipation regime.

When $Re$ << 1, molecular dissipation dominates. The absorption coefficient $\alpha_m$ for acoustic waves scales as (Sutherland and Bass, 2004):

$$\alpha_m \propto \nu\omega^2/c^3, \quad (3)$$

so that higher-frequency spectral components are attenuated quadratically faster than lower-frequency ones. In this regime, the atmosphere acts as a frequency-dependent low-pass filter: the high-frequency portion of the spectrum is irreversibly dissipated, the waveform is smoothed, and the surviving spectral content shifts toward progressively lower frequencies. The period measured at the ground then reflects the filter characteristics rather than, or in addition to, the source properties.

The attenuation term in Eq. (3) represents the classical thermoviscous molecular absorption used for the near-source scaling considered here. Vibrational relaxation can also contribute to atmospheric absorption, particularly at lower altitudes and for frequencies of a few hertz (Evans et al., 1972; Sutherland and Bass, 2004). In the present problem, the primary spectral pre-conditioning occurs within the first few kilometers below sources at 80–100 km, where the large kinematic viscosity makes thermoviscous attenuation the controlling effect. Relaxation effects during subsequent lower-atmosphere propagation may further modulate amplitudes and spectra, and are treated here as part of the broader path-propagation problem rather than the near-source filtering mechanism.

The transition between these regimes occurs over a narrow altitude range, and because $Re$ depends on frequency through $\omega_0$, the transition altitude shifts downward for higher-frequency signals. **Table 1** summarizes $Re$ for representative meteor infrasound parameters ($v_0 \approx 3$ m/s at $r_0$ = 100 m, $c$ = 300 m/s) at the two source frequencies $f_0$ = 2 Hz and 5 Hz. For a 2 Hz source, the formal $Re$ = 1 crossover occurs ~97-98 km; for a 5 Hz source, it shifts to below 94 km. However, significant spectral modification is already evident at $Re$ values of order 5-10 because the $\omega^2$ dependence of $\alpha_m$ ensures that higher-frequency spectral components are attenuated long before the fundamental is affected.

**Table 1:** Acoustic Reynolds number for representative meteor infrasound parameters ($v_0$ = 3 m/s, $c$ = 300 m/s) at two source frequencies. Kinematic viscosity $\nu$ values are from Gossard and Hooke (1975).

| Altitude (km) | $Re$ ($f_0$ = 2 Hz) | $Re$ ($f_0$ = 5 Hz) |
|---|---|---|
| 80 | ~55 | ~22 |
| 90 | ~5.5 | ~2.2 |
| 95 | ~1.7 | ~0.7 |
| 100 | ~0.55 | ~0.22 |
| $Re$ = 1 | ~97 – 98 km | ~93 – 94 km |

A critical feature of this dissipation is that it operates in the near-source zone, within the first few kilometers below the source, where the atmospheric density is still low and $\nu$ is large. Once the signal has descended into denser air (below ~70-75 km), $\nu$ drops rapidly and the effects of both nonlinearity and molecular dissipation become negligible. The signal then propagates essentially linearly to the ground, subject to refraction, geometric spreading, and scattering by fine-scale atmospheric layering, but without further significant spectral modification from molecular processes (Chunchuzov et al., 2025; Chunchuzov et al., 2026). This near-source confinement means that the spectral character of the signal arriving at the ground is largely determined by what happens in the first few km of propagation, not by the subsequent 70-100 km of descent to the receiver.

### 2.4 Implications for Regional Meteor Infrasound

The altitude dependence of $Re$ has immediate consequences for the interpretation of regional meteor infrasound. Small meteoroids frequently generate shocks at very high altitudes, often above 80 km and occasionally exceeding 100 km (Brown et al., 2007; Moreno-Ibáñez et al., 2018; Silber and Brown, 2014; Silber et al., 2026b). These altitudes place the acoustic source in or beyond the transitional dissipation regime. The signals from such events are therefore subject to near-source spectral filtering before they even begin their descent to the receiver.

This creates a specific, testable prediction. If near-source dissipation is the primary mechanism, one expects: (i) a monotonic decrease in dominant frequency with increasing source altitude, reflecting the altitude dependence of the filter cutoff; (ii) a progressive disappearance of high-frequency spectral content above ~80 km, where $Re$ drops into the dissipation-influenced range, so higher-frequency components are preferentially removed first; (iii) a convergence of dominant frequencies toward a narrow low-frequency band for sources above ~90–100 km, where only the lowest spectral components survive; and (iv) an independence of this spectral signature from propagation distance, because the filtering occurs near the source rather than along the extended path. All four predictions are amenable to direct observational test using a dataset with well-constrained source altitudes and receiver spectral characteristics, which the SOMN–ELFO regional meteor catalog provides.

From a practical standpoint, the implications are significant. Small meteoroids are by far the most numerous class of atmospheric entry events (Brown et al., 2002), and they constitute the majority of infrasound detections at regional networks. If the dominant periods of these abundant events are systematically lengthened by near-source dissipation, then period-yield relations calibrated on large bolides, whose low-altitude, low-frequency signals are minimally affected by this mechanism, will systematically overestimate the energies of small, high-altitude sources. The bias acts in one direction (longer periods means higher inferred energies) and grows with source altitude due to the exponential increase in viscosity. Quantifying this effect and defining the altitude regime in which period-based inference remains reliable is the primary objective of the present study.

The same physics applies to any high-altitude acoustic source, not only meteors. Space debris reentries, rocket body breakups, space mission returns, and other anthropogenic atmospheric entry events generate infrasound from comparable altitudes during entry or the early phases of breakup (e.g., Clemente et al., 2025; Hatty et al., 2026; Ishihara et al., 2012; Sansom et al., 2022; Silber et al., 2024). Infrasound-based characterization of these events is of growing interest for space situational awareness and atmospheric entry monitoring and faces the same near-source filtering constraints identified here.

## 3. Methods

### *3.1. Observational Data Set and Context*

This study uses the dataset of 71 distinct regional meteor events obtained by the Southern Ontario Meteor Network (SOMN) in combination with the Elginfield Infrasound Array (ELFO), Canada, as described in Silber and Brown (2014) and Silber et al. (2025a). The optical component consists of multi-station video and photographic observations that provide meteoroid trajectories, entry velocities, and photometric light curves. The acoustic component consists of infrasound waveforms recorded at ELFO. Several meteors produced more than one signal at the array, so that the 71 events generated a total of 90 infrasound detections.

From the full set of optically recorded meteors during the 2006–2011 interval, Silber and Brown (2014) identified those with coincident infrasound detections at ranges $R$ < 300 km. These regional infrasound meteors represent approximately 1% of the optically detected population and span a broad range of entry speeds (from slow asteroidal to fast cometary meteoroids) and entry geometries. For each event, optical astrometry yields a suite of parameters including the geocentric radiant, the entry velocity, the begin and end heights of luminous emission, and the meteor zenith angle. Photometry and ablation modeling further provide estimates of the pre-atmospheric mass for a subset of 24 events that are sufficiently bright and well constrained. The full dataset, including all optical solutions, infrasound recordings, and derived parameters, is publicly available at Zenodo (Silber et al., 2025b).

The infrasound analysis for each detection includes the dominant signal period, maximum and peak-to-peak amplitudes, signal celerity, back azimuth, and several quality and propagation metrics derived from ray tracing. While the present work focuses primarily on the period (and the independently estimated dominant frequency) and the raytraced best-fit source height, as these are most directly linked to the altitude-dependent filtering described in **Section 2.3**, the richer set of infrasound parameters in the underlying catalog allows assessment of signal quality, propagation complexity, and consistency across stations. Where necessary, these additional parameters are consulted to exclude ambiguous detections and to confirm the robustness of the event classifications.

The geographic context of the dataset is illustrated in **Figure 2**, which shows the projected ground tracks of all meteoroids together with the locations of the raytraced infrasound source points. The meteors were observed over southern Ontario and the adjacent Great Lakes region, with ELFO located near the center of the network. Each source point is plotted as a circle colored by its inferred shock source height, and grey line segments indicate the corresponding luminous trajectories derived from optical astrometry. The figure demonstrates that all sources lie at regional ranges (< 300 km) from ELFO and that the infrasound-generating segments of the trajectories are distributed broadly around the array, sampling source heights from approximately 20 to just over 100 km.

In all subsequent analyses, each infrasound detection is treated as an individual data point. Distinct arrivals from the same meteoroid can correspond to different effective source points along the trajectory and, in some cases, to different propagation paths through the atmosphere. Treating detections independently preserves this physical distinction.

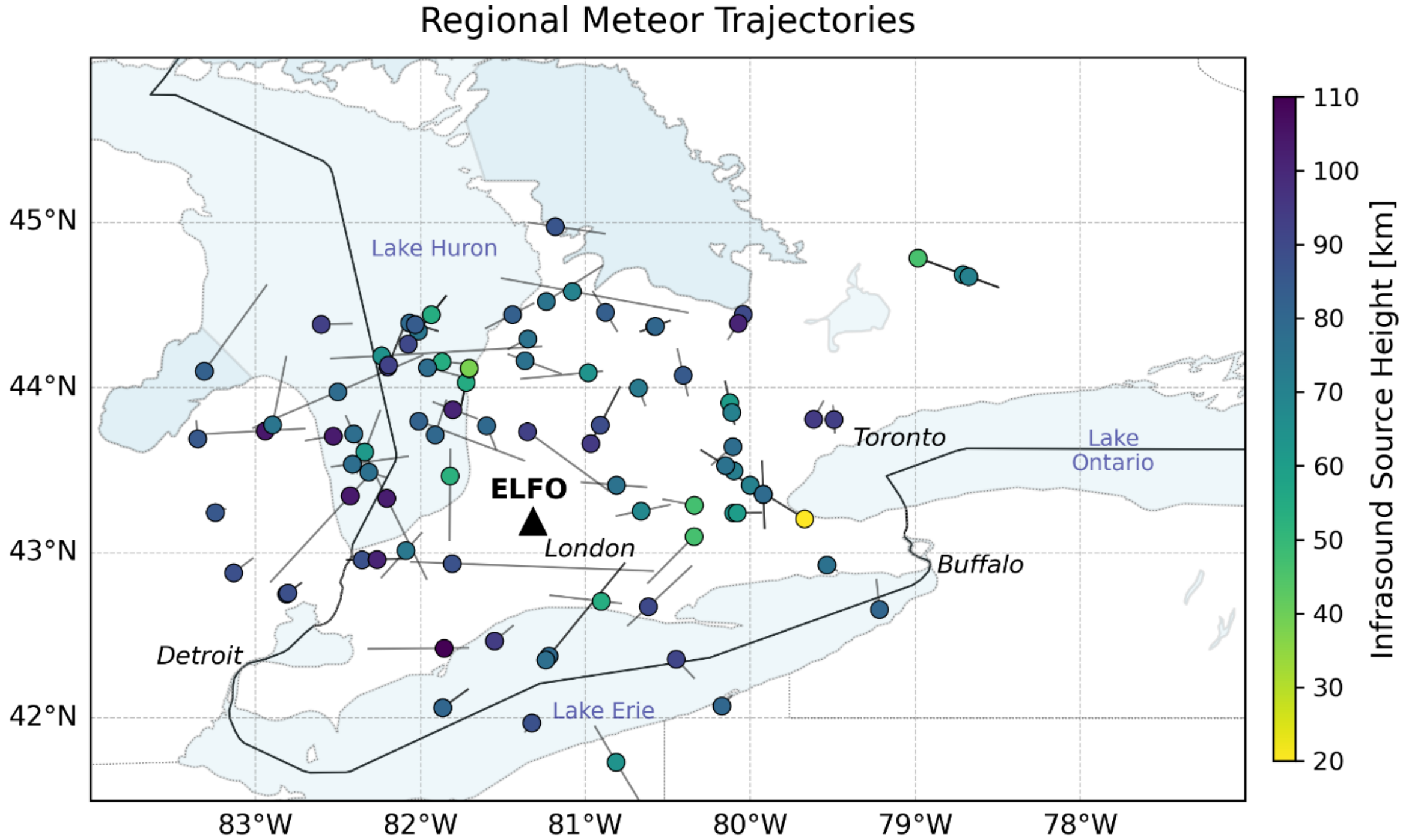

**Figure 2:** Map of southern Ontario and the Great Lakes region showing the locations of the 71 meteoroid trajectories relative to the Elginfield Infrasound Array (ELFO, black triangle). Circles mark the projected ground positions of the raytraced infrasound source points, colored by source height. These source heights were obtained by raytracing the infrasound arrivals to the independently determined optical meteor trajectories. Grey line segments indicate the projected ground tracks of the meteors derived from optical astrometry. All events are detected at regional ranges (< 300 km) and the infrasound sources sample altitudes from ~20 to 111 km.

### 3.2 Statistical Analysis

The primary observational analysis tests for systematic relationships between source altitude and receiver spectral content. We employ rank-based (Spearman) correlations throughout because they are robust to outliers and do not assume linearity; the altitude–period and altitude–frequency relationships span more than an order of magnitude in spectral content and exhibit potential nonlinearity that could bias parametric (Pearson) statistics.

Because source altitude, propagation distance, and entry velocity are potentially confounded (e.g., faster meteoroids ablate at higher altitudes, and higher sources are necessarily farther from the receiver), we compute rank-based partial correlations to isolate the altitude effect. Partial correlations remove the linear dependence of both variables on a third (the confound) by correlating the residuals of rank regressions against the confounding variable. We control separately for horizontal range, three-dimensional source–receiver range, and entry velocity to evaluate whether the observed spectral trends are attributable to propagation distance, to the meteoroid population characteristics, or to the atmospheric state at the source.

Detections are grouped into four altitude bins (below 60 km, 60–80 km, 80–100 km, and above 100 km) for distributional comparison. The below-60-km bin represents the regime where $Re$ >> 1 and spectral content is expected to be preserved (in case of small Mach number $M_0$=$v_0/c$<<*1* when a nonlinear stretching of N-wave is small); the 80–100 km bin spans the transitional zone; and the above-100-km bin captures the fully dissipation-dominated regime. The 60–80 km bin provides intermediate coverage. Within each bin we compute means, standard deviations, and coefficients of variation for both period and frequency. One-sided Mann–Whitney U tests are used to compare the period distributions above and below candidate altitude thresholds (75, 80, 85, and 90 km) without assuming normality.

To characterize the progressive extinction of high-frequency content with altitude, we construct frequency survival curves: for each of several frequency thresholds (1, 2, 3, and 5 Hz), we compute the fraction of detections with dominant frequencies exceeding that threshold as a function of the minimum source altitude.

### *3.3. Near-Source Dissipation Modeling*

To connect the observed spectral trends to the physical mechanism described in **Section 2.3**, we model the near-source transformation of N-wave signals from point sources at altitudes of 80, 90, and 100 km. The modeling follows the generalized Burgers equation framework for spherically diverging waves in a stratified atmosphere (Chunchuzov et al., 2013; Kshevetskii et al., 2024; Ostrovskii et al., 1976; Rogers and Gardner, 1980; Rudenko and Soluyan, 1977; Sabatini et al., 2016).

The fragmentation of a meteoroid is modeled as a point source of an acoustic N-wave at altitude $z = z_1$. The radial velocity field $v(r,t)$ is taken to diverge spherically from the source, with the dimensionless velocity defined as $u(r,t) = \frac{v(r,t)}{v_0(r_0,t_0)}$, where $v_0$ is the peak velocity amplitude at a reference distance $r_0$ = 100 m. Ray tracing calculations for sources above 80 km (Chunchuzov et al., 2025, Figures 2–5) confirm that refraction weakly affects the spherically diverging ray paths out to distances of several kilometers, justifying the spherical divergence assumption over the near-source propagation interval.

The initial conditions are specified with a Mach number $M_0 = \frac{v_0}{c} = \frac{P_0'}{\rho_a c^2} = 0.01$ at $r_0$ = 100 m, where $v_0$ is the peak particle velocity of the N-wave disturbance at $r_0$, $P_0'$ is the corresponding peak pressure disturbance amplitude, $\rho_a$ is the mean atmospheric density, and $c$ is the local sound speed. We use $v_0$ = 3 m/s, a representative value for a finite-amplitude acoustic disturbance from a small meteoroid fragmentation in the rarefied upper atmosphere. We note that $M_0$ is the Mach number of the acoustic wave, not the hypersonic flight Mach number of the meteoroid; at $r_0$ = 100 m from a fragmentation point, the shock has decayed well below the near-field hypersonic regime.

The reference amplitude is set to $v_0 = 3$ m/s at $r_0 = 100$ m, corresponding to an acoustic Mach number $M_0 \approx 0.01$ for $c \approx 300$ m/s. This value is adopted as a representative weakly nonlinear N-wave amplitude following the finite-amplitude acoustic scaling used in nonlinear-acoustic and thermospheric propagation treatments (Chunchuzov et al., 2013; Rogers and Gardner, 1980;

Rudenko and Soluyan, 1977; Sabatini et al., 2016). It should be interpreted as a reference amplitude for the modeled acoustic disturbance after the initial meteor shock has decayed from the immediate hypersonic near field. The adopted source frequencies $f_0 = 2$ and 5 Hz were chosen to represent a typical regional-meteor dominant frequency and a higher-frequency case within the observed SOMN–ELFO range (Silber and Brown, 2014). Event-to-event values are expected to vary with source energy and fragmentation geometry; a representative sensitivity range of $v_0 \sim 1$–10 m/s would shift $Re$ by the same factor. Because $Re$ scales linearly with $v_0$, while kinematic viscosity varies by orders of magnitude over the 80–100 km altitude interval, such variations shift the $Re \sim 1$ transition altitude by only several kilometers and leave the qualitative spectral-filtering behavior unchanged.

Two representative initial signal durations are considered:

- Case 1: $T_0$ = 0.5 s (dominant frequency $f_0$ = 2 Hz), evaluated at source altitudes of 80 km (Case 1a), 90 km (Case 1b), and 100 km (Case 1c). This initial frequency represents the typical low-frequency content commonly recorded at the ground for regional meteor events.
- Case 2: $T_0$ = 0.2 s ($f_0$ = 5 Hz), evaluated at 90 km. This higher-frequency case addresses the spectral content characteristic of smaller meteoroids and tests the survival limits of spectral components above 3 Hz.

The choice of a point-source geometry warrants comment. Meteoroid entry generates shock waves through two distinct mechanisms: a quasi-cylindrical ballistic shock along the trajectory, and localized spherical blasts at discrete fragmentation points along the path. These descriptions correspond to idealized source representations commonly used in meteor-infrasound interpretation. The ballistic component is often treated using cylindrical or line-source weak-shock theory along the trajectory (e.g., Plooster, 1970; ReVelle, 1976; Silber et al., 2015; Tsikulin, 1970). Fragmentation-associated arrivals are commonly represented as effective localized acoustic sources from finite regions of rapid energy deposition, especially in propagation studies where the source location is constrained independently (e.g., Chunchuzov et al., 2025; Edwards et al., 2006; Scamfer et al., 2026; Silber et al., 2018). A fully predictive model linking detailed fragmentation physics to the initial infrasound source spectrum remains an open problem. The point-source treatment used here is therefore an idealized acoustic representation chosen to isolate the altitude-dependent near-source filtering of the emitted spectrum. In the SOMN-ELFO dataset, both source types are represented; the raytraced source points include both ablation-generated cylindrical shocks and discrete fragmentation episodes. The point-source modeling presented here is directly applicable to the fragmentation events, which produce quasi-spherical blast waves from localized energy releases.

For the cylindrical line-source component, the geometry differs in its amplitude decay rate, approximately $r^{-1/2}$ for ideal cylindrical spreading compared with $r^{-1}$ for ideal spherical spreading, and in the detailed form of the nonlinear waveform evolution. The slower geometrical decay of a cylindrical disturbance can allow finite-amplitude effects to persist over greater distances, so the quantitative balance between nonlinear steepening, nonlinear stretching, and thermoviscous attenuation may differ from the spherical point-source calculation presented here (e.g., ReVelle, 1976; Silber et al., 2018). The acoustic Reynolds number $Re$ remains a useful local nondimensional measure of the competition between finite-amplitude nonlinear evolution and molecular dissipation for a waveform propagating at a given altitude. However, the altitude-dependent filtering emphasized here is governed primarily by the local atmospheric state and signal frequency through the molecular absorption coefficient $\alpha_m$ (Eq. (3)). Spectral components generated by either a localized quasi-spherical disturbance or a cylindrical ballistic disturbance traverse the high-viscosity near-source environment, where higher frequencies are preferentially attenuated. The point-source calculation is therefore used here as a tractable first-order representation of near-source spectral pre-conditioning, while quantitative cylindrical-wave calculations are identified as a useful extension of this work.

The observational analysis (**Section 4.1**), which examines the full dataset without distinguishing between source types, provides a direct empirical test of the altitude-frequency relationship that is independent of any modeling assumptions about source geometry. The statistical trends reflect what the atmosphere delivers to the receiver, regardless of the initial shock configuration. The modeling then provides a physical explanation for those trends using the analytically tractable point-source case. From a practical standpoint, the mode of shock production for a given meteoroid event is often not known with certainty, particularly for faint events detected only by infrasound without detailed optical fragmentation data. This work therefore presents an important consideration for interpreting dominant periods from any high-altitude infrasound source, regardless of the assumed or inferred shock geometry.

These cases span the parameter space identified in **Section 2.3** as critical: Case 1a (80 km, $Re \sim 55$) lies firmly in the weakly nonlinear regime, while Cases 1b–1c (90–100 km, $Re$ ~ 5.5-0.55) are in the transitional and dissipation-dominated regime. Case 2 tests whether higher-frequency sources fare worse, as predicted by the $\omega^2$ dependence of the absorption coefficient.

For the weakly nonlinear regime ($Re$ >> 1), the signal evolution is computed using the analytical expressions for nonlinear N-wave stretching, amplitude reduction, and shock-front smoothing given by Rudenko and Soluyan (1977) (see **Section 4.3.1**). For the dissipation-dominated regime ($Re$ << 1), the frequency spectrum of the N-wave at distance $r$ from the source is obtained by multiplying its near-source spectrum (at $r_0$ = 100 m) by the dissipation factor $\exp[-\alpha_m(\omega)(r - r_0)]$, where $\alpha_m(\omega)$ is the molecular absorption coefficient (Eq. (3)), and applying an inverse Fourier transform to recover the time-domain signal at $r$ = 5 km.

The choice of $r$ = 5 km as the evaluation distance is motivated by two considerations. First, over this distance the high-altitude thermoviscous loss considered here has already imposed substantial spectral pre-conditioning on the signal. Second, at ranges $r$ > 5 km, further propagation of the signal with its surviving low-frequency spectrum is affected mostly by refraction and scattering of its spectral components by the fine-scale layered structure of the atmosphere crossing the ray paths (Chunchuzov et al., 2025; Chunchuzov et al., 2026). The 5 km distance is therefore used as a practical reference distance for isolating the near-source filtering mechanism, not as a sharp physical boundary beyond which spectral evolution ceases. Parabolic equation calculations show that scattering does not significantly affect the direct arrival of the signal on the ground up to horizontal ranges of 200 km from the source (Chunchuzov et al., 2025, Figures 2-4). The 5 km distance therefore represents the boundary between near-source spectral pre-conditioning and far-field linear propagation.

**4. Results and Discussion**

The SOMN–ELFO regional meteor sample encompasses 71 distinct meteoroid events producing 90 infrasound detections at horizontal ranges of 47–268 km (121.3 ± 43.3 km) from the Elginfield array. The distributions of principal observational and derived parameters are summarized in **Figure 3**.

Raytraced effective source altitudes span 20-111 km, with a mean of 77.8 ± 16.2 km and a pronounced concentration between 70 and 95 km; 11 detections originate below 60 km and 8 above 100 km. Entry velocities range from 11.5 to 75.6 km/s (38.4 ± 19.2 km/s), reflecting the full spectrum from slow asteroidal to fast cometary meteoroids. Entry angles (radiant altitude) are broadly distributed (45.2 ± 16.2 degrees) and are not significantly correlated with source altitude ($r_s$ = -0.131, $p$ = 0.22). Luminous begin heights average 97.8 ± 14.0 km and end heights average 65.4 ± 18.7 km, indicating that the acoustic source regions sampled by ELFO are located within or near the upper portions of the luminous trajectories for the majority of events. Total three-dimensional source-receiver ranges average 146.5 ± 37.9 km (73-277 km).

Receiver dominant periods span 0.07-2.56 s, with a right-skewed distribution (mean 0.64 ± 0.46 s; median 0.53 s) and a mode near 0.3-0.5 s. The corresponding dominant frequencies range from 0.4 to 12.9 Hz (mean 2.5 ± 2.3 Hz; median 1.9 Hz), consistent with the sub-20 Hz infrasonic regime expected for small regional meteor sources. Both the period and frequency distributions (**Figure 3b,c**) display marked right-skewed character, with the frequency histogram peaking sharply in the 0.5-1.5 Hz band. These distributions establish that the sample is dominated by high-altitude sources whose ground-recorded spectral content falls predominantly in the 1-3 Hz band, with a substantial tail toward both higher and lower frequencies.

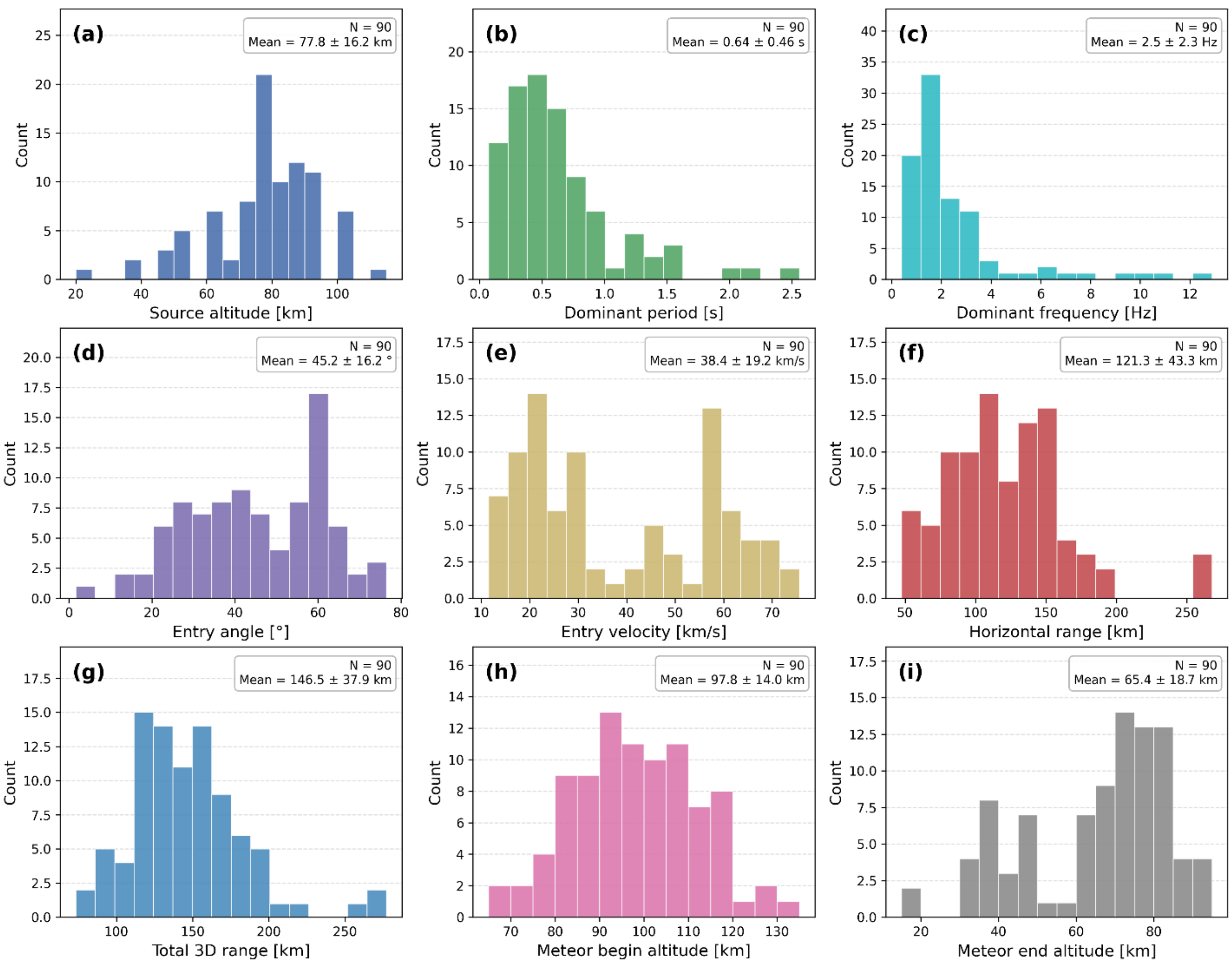

**Figure 3:** Summary distributions for the regional meteoroid infrasound dataset ($N$ = 90 detections from 71 events). Panels show: (a) effective acoustic source altitude, (b) receiver dominant period, (c) receiver dominant frequency, (d) entry angle (radiant altitude), (e) entry velocity, (f) horizontal source-receiver range, (g) total three-dimensional source-receiver range, (h) meteor luminous start altitude, and (i) meteor luminous end altitude. Inset annotations report the sample size, mean, and standard deviation for each parameter.

### 4.1 Altitude Dependence of Receiver Spectral Content

The primary observational finding of this study is a strong, monotonic relationship between effective source altitude and the spectral characteristics of the ground-recorded infrasound. **Figure 4** presents source altitude plotted against dominant period (panel a) and dominant frequency (panel b), both on logarithmic horizontal axes. The data reveal an evident positive trend between altitude and period and a corresponding negative trend between altitude and frequency: higher-altitude sources produce systematically longer-period, lower-frequency ground signals.

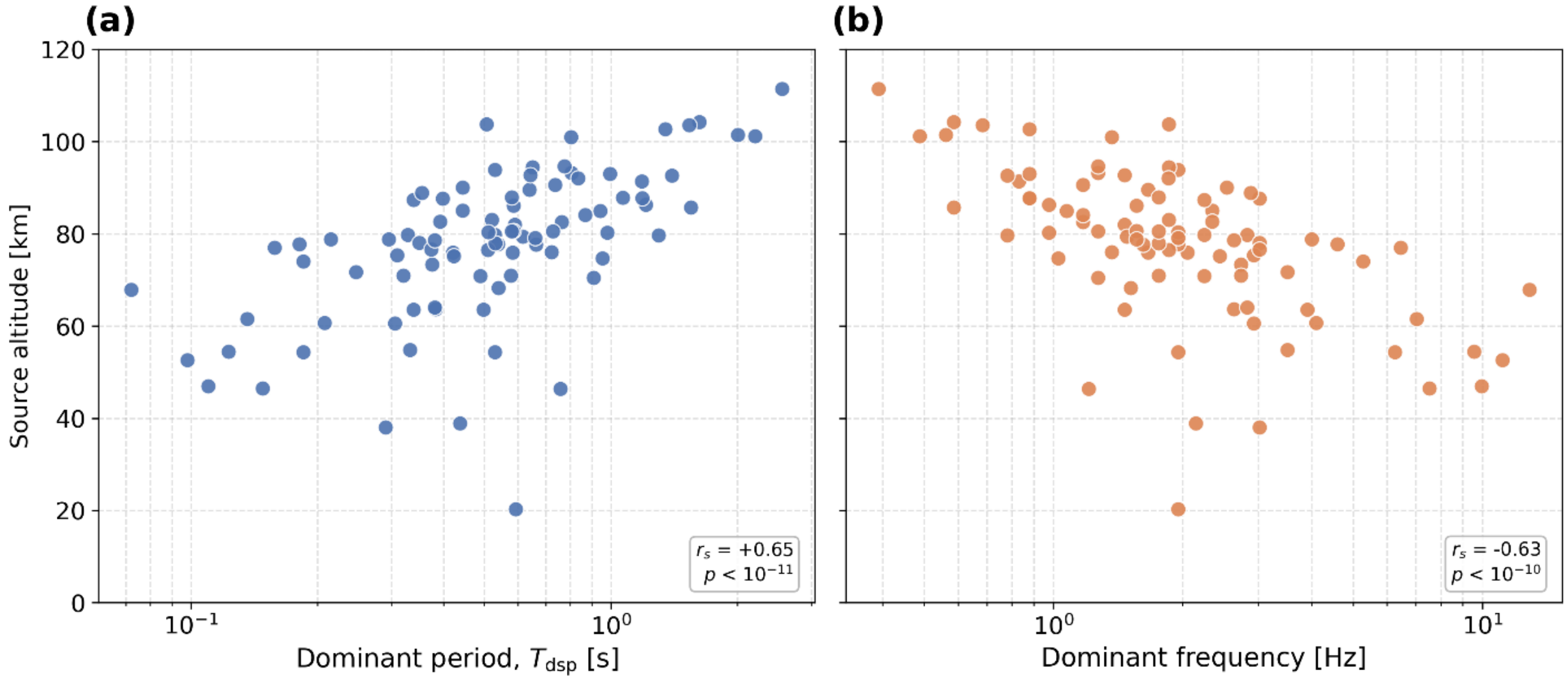

**Figure 4:** Source altitude versus dominant period (a) and dominant frequency (b), both on logarithmic horizontal axes.

The altitude-period correlation is statistically robust. Spearman rank correlation yields $r_s$ = +0.645 ($p$ = 6.7 x $10^{-12}$), indicating a strong monotonic association. The altitude-frequency anticorrelation is comparably strong ($r_s$ = -0.629, $p$ = 3.3 x $10^{-11}$). Both correlations are somewhat stronger in log-transformed space (Pearson $r_p$ = +0.600 for $\log_{10}$(period), $r_p$ = -0.596 for $\log_{10}$(frequency)), consistent with an approximately log-linear relationship between source height and spectral content.

Because source altitude, propagation distance, and entry velocity are potentially confounded (faster meteoroids tend to generate shocks at higher altitudes, $r_s$ = +0.705, $p$ = 9.1 x $10^{-15}$, and longer propagation paths provide more opportunity for frequency-dependent attenuation), we computed rank-based partial correlations to isolate the altitude effect (**Section 3.2**). Controlling for horizontal range, the altitude-period association remains essentially unchanged (partial $r_s$ = +0.650), and controlling for three-dimensional range yields a similarly robust result (partial $r_s$ = +0.558). These values confirm that the spectral trend is primarily an altitude effect rather than a propagation distance artifact. Controlling for entry velocity reduces the partial correlation to $r_s$ = +0.360 ($p$ = 5.0 x $10^{-4}$), indicating that while velocity contributes to the observed trend through its influence on source altitude, the altitude-period relationship retains substantial independent explanatory power beyond what velocity alone can account for.

The progressive spectral shift with altitude is quantified in **Figure 5**, which presents the distributions of dominant period (panel a) and dominant frequency (panel b) across the four altitude bins defined in **Section 3.2**. Detections originating below 60 km ($N$ = 11) exhibit the broadest spectral content, with frequencies spanning 1.2-11.1 Hz (mean 5.3 ± 3.7 Hz) and periods of 0.10-0.76 s (mean 0.33 ± 0.22 s). In the 60-80 km range ($N$ = 38), mean frequency decreases to 2.9 ± 2.2 Hz and mean period increases to 0.45 ± 0.24 s. Between 80 and 100 km ($N$ = 33), the distributions narrow markedly: mean frequency drops to 1.6 ± 0.6 Hz and mean period rises to 0.75 ± 0.31 s, with the frequency coefficient of variation decreasing from 0.70-0.73 in the lower bins to 0.39 in this range. For the eight detections above 100 km, mean frequency is 0.9 ± 0.5 Hz and mean period reaches 1.57 ± 0.69 s; six of the eight (75%) have dominant frequencies below 1 Hz. The violin distributions in **Figure 5** make two features visually apparent: the systematic shift in central tendency across bins and the collapse of the upper spectral tail at higher altitudes. The broad, asymmetric distributions seen below 60 km progressively compress into a narrow low-frequency band above 80 km.

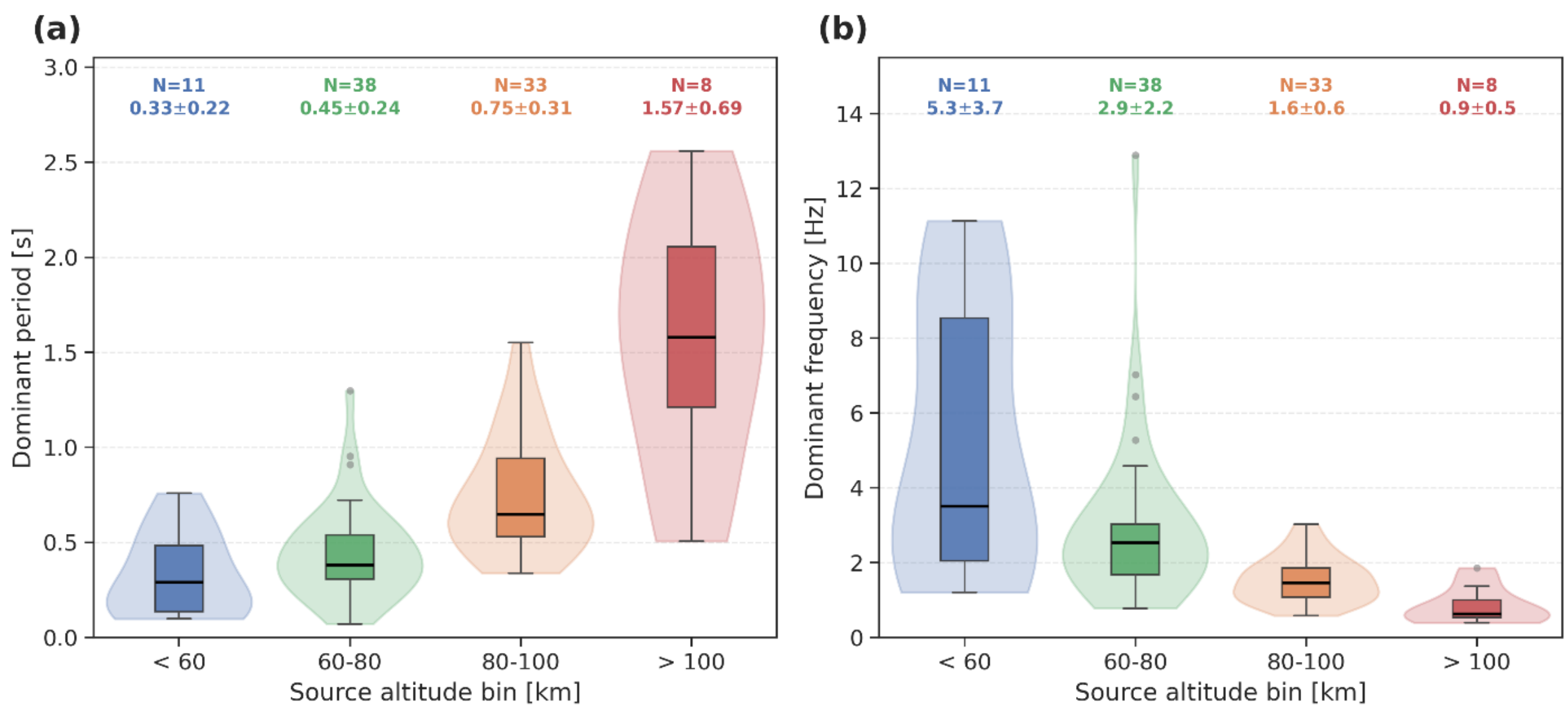

**Figure 5:** Distributions of dominant period (a) and dominant frequency (b) across four altitude bins: below 60 km ($N$ = 11), 60-80 km ($N$ = 38), 80-100 km ($N$ = 33), and above 100 km ($N$ = 8). Violin outlines show the kernel density estimate of each distribution; embedded box plots indicate the median (white circle), interquartile range (thick bar), and 1.5x interquartile range (thin line). The progressive shift in central tendency and the collapse of the upper spectral tail with increasing altitude are both apparent.

A Mann-Whitney U test confirms that the period distributions above and below 80 km differ significantly ($U$ = 1706, $p$ = 6.8 x $10^{-9}$), with the above-80-km population shifted to periods approximately twice as long (mean ratio 2.2x; median ratio 2.0x). The significance holds across alternative thresholds: at 75 km ($p$ = 3.3 x $10^{-6}$, ratio 2.0x), 85 km ($p$ = 1.4 x $10^{-7}$, ratio 2.2x), and 90 km (p = 1.1 x $10^{-6}$, ratio 2.2x). The altitude dependence is expressed most clearly through the systematic disappearance of high-frequency spectral content with increasing source height. **Figure 6** presents the fraction of detections retaining dominant frequencies above 1, 2, 3, and 5 Hz as a function of minimum source altitude threshold.

Among detections with sources below 80 km, 63% (31 of 49) exhibit dominant frequencies exceeding 2 Hz, and 31% (15 of 49) exceed 3 Hz. Above 80 km, only 15% (6 of 41) exceed 2 Hz, and none exceeds 3 Hz. Above 90 km, a single detection out of 19 (5%) retains a dominant frequency above 2 Hz, and no detections above 80 km have frequencies exceeding 5 Hz. The $f$ > 3 Hz survival curve drops to zero at 80 km, the $f$ > 5 Hz curve at ~78 km, and the $f$ > 2 Hz curve falls from ~40% to effectively zero between 80 and 100 km. Conversely, the fraction of detections dominated by sub-1 Hz content rises from 16% for sources above 60 km to 29% above 80 km, 47% above 90 km, and 75% above 100 km.

These results support the notion that the atmosphere imposes an altitude-dependent spectral filter on regional meteor infrasound, as anticipated by the theoretical framework in **Section 2.3**. High-frequency content is progressively and systematically extinguished as source altitude increases, with near-complete suppression of spectral components above 2-3 Hz for sources in the upper mesosphere and lower thermosphere. This behavior is not attributable to differences in source-receiver range, is only partially accounted for by entry velocity, and points to a physical mechanism operating in or near the source region.

The altitude dependence of the received spectral content is also evident in representative waveform-derived spectra. **Figure 7** shows a recorded amplitude spectrum for one detection in each altitude bin, computed from the ELFO array data by windowing each signal at its catalogued arrival time and duration, applying the per-event band-pass filter, and averaging the power spectral density across the four array elements. The examples illustrate the same progression quantified statistically in **Figures 4–6**: the sub-60 km case retains broadband energy extending beyond 10 Hz; the 60–80 km and 80–100 km cases are increasingly concentrated below a few hertz; and the >100 km case is dominated by sub-1 Hz content with little energy above ~2 Hz. This waveform-derived spectral view reinforces the statistical result that high-frequency spectral content is progressively reduced as source altitude increases.

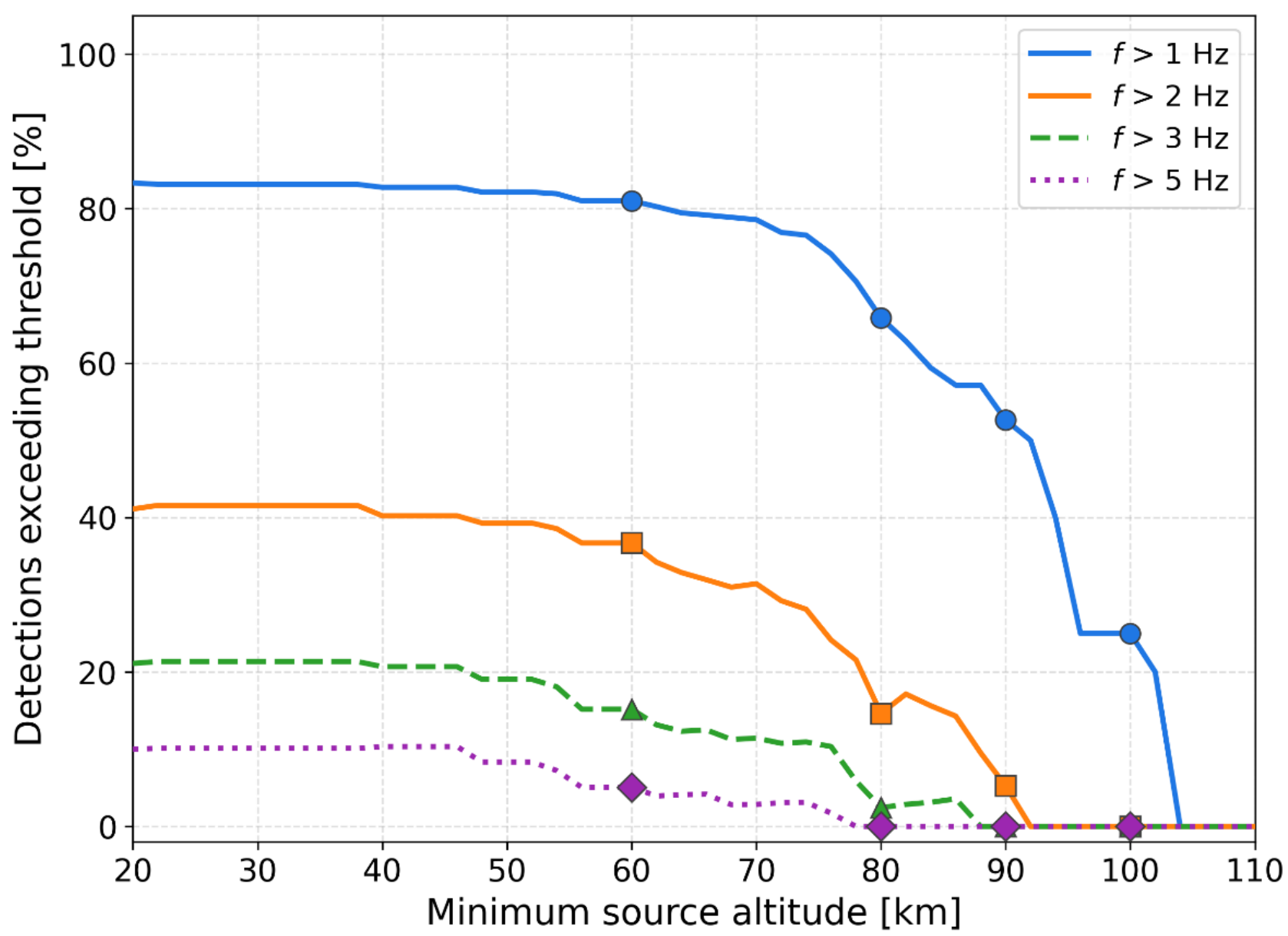

**Figure 6:** Fraction of detections with dominant frequency exceeding threshold values of 1 Hz (blue), 2 Hz (orange), 3 Hz (green), and 5 Hz (purple) as a function of the minimum source altitude. The $f$ > 3 Hz fraction drops to zero at 80 km, and the $f$ > 2 Hz fraction approaches zero above 90 km.

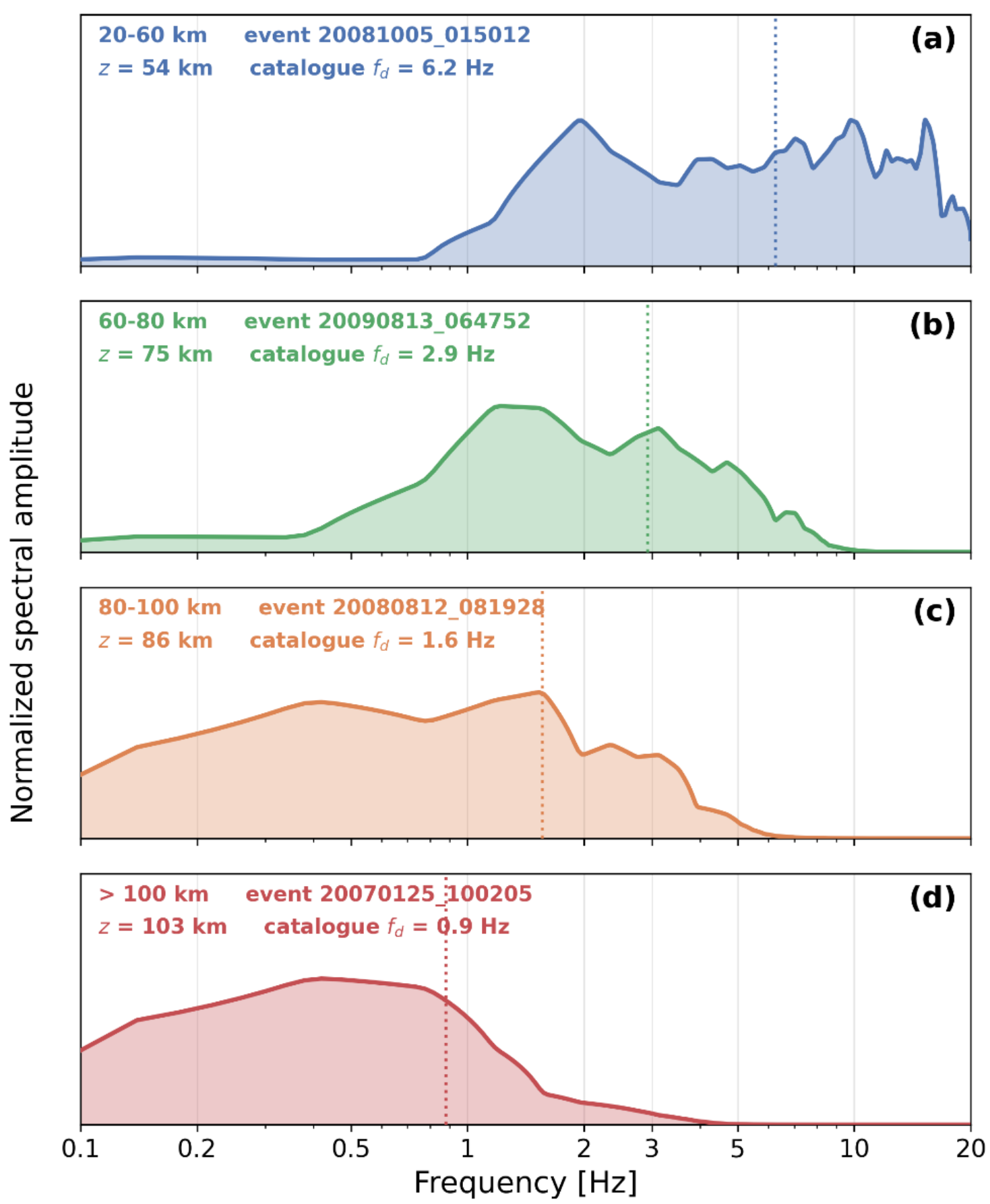

**Figure 7:** Representative recorded infrasound amplitude spectra as a function of source altitude. Each panel shows one detection from an altitude bin, selected as the detection whose catalogued dominant frequency is closest to the bin mean: (a) 20–60 km, event 20081005_015012, $z = 54$ km; (b) 60–80 km, event 20090813_064752, $z = 75$ km; (c) 80–100 km, event 20080812_081928, $z = 86$ km; and (d) >100 km, event 20070125_100205, $z = 103$km. Spectra were computed directly from the ELFO array waveforms. Each signal was windowed at its catalogued arrival time and duration, band-pass filtered using the per-event corner frequencies, and the power spectral density was averaged over the four array elements (ELFO1–4). The resulting amplitude spectrum is normalized to its own maximum. The dotted vertical line marks the catalogued dominant frequency $f_d$, which is measured from the zero-crossing dominant period (Silber et al., 2025a; Silber and Brown, 2014) and therefore need not coincide exactly with the spectral peak. These representative spectra illustrate the progressive shift of recorded spectral content toward lower frequencies with increasing source altitude.

### 4.2 Near-Source Dissipation Physics

The observational trends are now compared against the near-source dissipation modeling. The modeling evaluates the transformation of N-wave signals from point sources at 80, 90, and 100 km using the generalized Burgers equation framework, with initial conditions as specified in **Section 3.3** ($M_0$ = 0.01 at $r_0$ = 100 m; $T_0$ = 0.5 s and 0.2 s) (Chunchuzov et al., 2013; Ostrovskii et al., 1976; Rogers and Gardner, 1980; Rudenko and Soluyan, 1977; Sabatini et al., 2016).

**4.2.1 Case 1a: Weakly Nonlinear Regime: $f_0$ = 2 Hz at 80 km ($Re$ ~ 55)**

At 80 km, where $Re$ ~ 55, the N-wave undergoes two competing processes as it propagates outward from the source. Nonlinear wave propagation leads to a stretching of the N-wave duration $T(r)$ with increasing distance $r$ from the source (Rudenko and Soluyan, 1977):

$$T(r) = T_0 [1 + s(r)]^{1/2}, \quad (4)$$

where

$$s(r) = (r_0/x_{non}) \ln(r/r_0), \quad (5)$$

and $x_{non} = [\frac{(\gamma+1)}{2}\frac{\omega_0}{c}M_0]^{-1}$ is the characteristic length at which a discontinuity arises in an initially plane harmonic wave of frequency $\omega_0$ propagating in an ideal (inviscid) atmosphere, with $\gamma$ being the ratio of specific heats (the adiabatic constant). The nonlinear stretching of the positive and negative phases of the N-wave is accompanied by a decrease in the peak velocity amplitude with increasing distance $r$:

$$v_{peak}(r) = v_0\left(\frac{r_0}{r}\right)[1 + s(r)]^{-1/2}, \quad (6)$$

accounting for both spherical divergence and nonlinear amplitude reduction.

At the same time, molecular absorption "smooths" the shock front, so that the thickness of the shock front, $\tau_{shock}(r)$, increases with distance $r$ from the source. If the radial velocity $v(r,t)$ in a spherical wave is specified at $r = r_0$ as a step jump from $-v_0$ to $v_0$ at the initial time $t = 0$, then after propagating to distance $r$ the jump is "smoothed" and takes the following dimensionless form (Rudenko and Soluyan, 1977):

$$u(r,t) = \frac{v(r,t)}{\left(\frac{v_0\ r_0}{r}\right)} = \tanh\left(\frac{t_{sig}}{\tau_{shock}}\right), \quad (7)$$

where $t_{sig} = t - \frac{(r - r_0)}{c}$ is the time in the reference frame accompanying the signal. The shock front thickness $\tau_s(r)$ is:

$$\tau_{shock} = \nu\frac{\left(\frac{r}{r_0}\right)}{\left[\left(\frac{\gamma+1}{2}\right)v_0 c\right]} = T_0\frac{\left(\frac{r}{r_0}\right)}{\left[2\pi\left(\frac{\gamma+1}{2}\right)Re\right]}. \quad (8)$$

The two expressions for $\tau_{shock}$ in Eq. (8) are equivalent: the first relates the shock thickness directly to the ratio of propagation distance to the viscous length scale $\left(\frac{v_0}{\nu}\right)^{-1}$, while the second expresses the same quantity in terms of the initial signal duration and the acoustic Reynolds number.

With the N-wave parameters and $Re$ ~ 55 specified for Case 1a, the shock front thickness at a distance of $r$ ~ 5 km is only ~0.1 $T_0$. The duration $T$ of the N-wave increases in accordance with Eq. (4)-(5) by only a factor of ~1.1, and its amplitude (Eq. (6)) relative to the amplitude of the purely spherically diverging wave decreases by a factor of ~1.1 (or a factor of 55 in absolute terms, taking into account the spherical divergence of the wave). As a result, the N-wave, defined at $r = r_0$ in dimensionless form $u(r,t) = \frac{v(r,t)}{\left(\frac{v_0\ r_0}{r}\right)}$ (Eq.(7)) for $-T_0/2$ < t < $T_0/2$ (**Figure 8**, upper panel, dotted line), acquires the form $u\left(r, t_{sig}\right)$ at $r$ = 5 km shown in **Figure 8** (upper panel, 80 km curve). This form does not differ significantly from the initial N-wave profile, indicating a weak influence of nonlinearity and dissipation on the waveform and Fourier spectrum as the signal propagates to this distance (**Figure 8**, lower panel, 80 km curve). The main effect is a slight shift of the dominant frequency to the low-frequency region and a decrease in the amplitudes of higher-frequency spectral maxima. The spectrum preserves significant content up to at least 2 Hz.

This result is consistent with the observations: sources below ~80 km in the SOMN-ELFO dataset retain broad spectral content, with 63% exhibiting dominant frequencies above 2 Hz and 31% above 3 Hz. As the signal propagates farther from $r$ = 5 km toward the ground, the effects of nonlinearity and dissipation rapidly diminish due to the exponential increase in mean atmospheric density with decreasing altitude. The subsequent waveform is therefore expected to be influenced mainly by linear propagation processes and scattering by fine-scale effective-sound-speed structure, while additional path and measurement effects may still contribute to the observed spectral shape (Chunchuzov et al., 2025).

#### 4.2.2 Cases 1b-1c: Transitional and Dissipation-Dominated Regimes, Sources at 90-100 km

At altitudes of 90-100 km, the kinematic viscosity increases rapidly, reducing $Re$ from ~5.5 at 90 km to ~0.55 at 100 km (**Table 1**). At 100 km, Re drops below unity and molecular absorption fully prevails over nonlinear effects, allowing the problem to be treated within a linear approximation (Rudenko and Soluyan, 1977). As a first approximation, for both altitudes the frequency spectrum of the N-wave at distance $r$ is obtained by multiplying its near-source spectrum (at $r_0$ = 100 m) by the dissipation factor exp$[-\alpha_m(\omega)(r - r_0)]$, and applying an inverse Fourier transform.

Case 1b: $f_0$ = 2 Hz at 90 km ($\nu$ ~ 13 m$^2$/s). The absorption of high-frequency spectral components at $r$ = 5 km leads to strong attenuation of components with frequencies above ~4 Hz (**Figure 8**, lower panel, solid 90 km curve) and significant smoothing of the signal profile (**Figure 8**, upper panel, 90 km curve). The dominant frequency shifts downward from the initial 2 Hz. This is consistent with the observations: in the 80-100 km altitude bin, no detections retain dominant frequencies above 3 Hz, and the mean frequency is 1.6 ± 0.6 Hz.

Case 1c: $f_0$ = 2 Hz at 100 km ($\nu$ ~ 130 m$^2$/s). The tenfold further increase in viscosity produces strong attenuation of the high-frequency portion of the emitted signal spectrum, shifting its main maximum from $f_0$ = 2 Hz down to ~1 Hz at $r$ = 5 km (**Figure 9**). Spectral components with frequencies > 2 Hz are strongly absorbed, and the signal duration at this distance is approximately twice the initial N-wave duration. This maps directly onto the observations: the eight detections >100 km have a mean dominant frequency of 0.9 ± 0.5 Hz, with 75% dominated by sub-1 Hz content. With the increase in fragmentation height up to 100 km, dominant frequencies detected on the ground are predominantly <1 Hz (**Figures 6 and 7**), which is explained by the shift of the dominant frequency to the low-frequency region due to the strong attenuation of spectral components above 2 Hz.

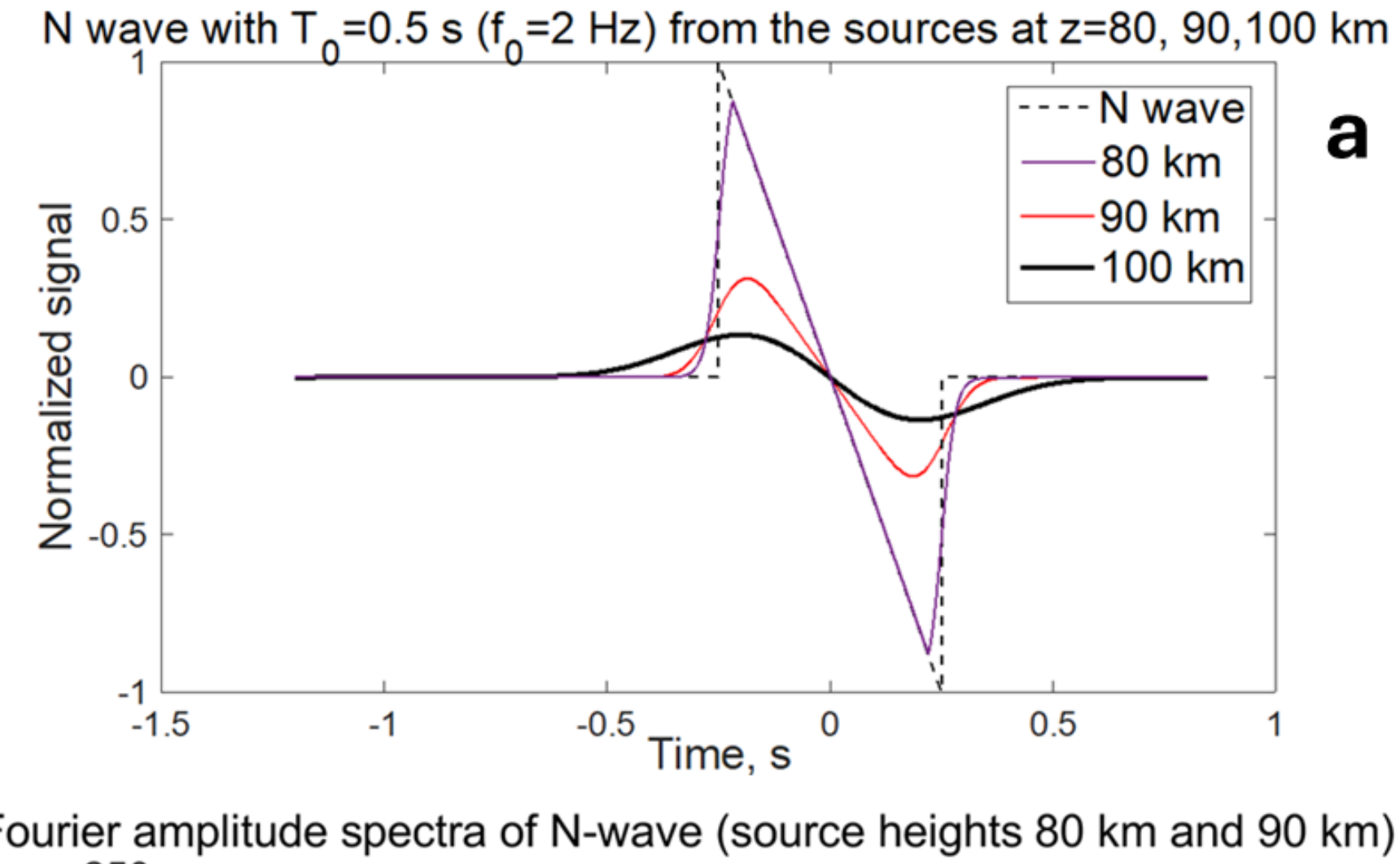

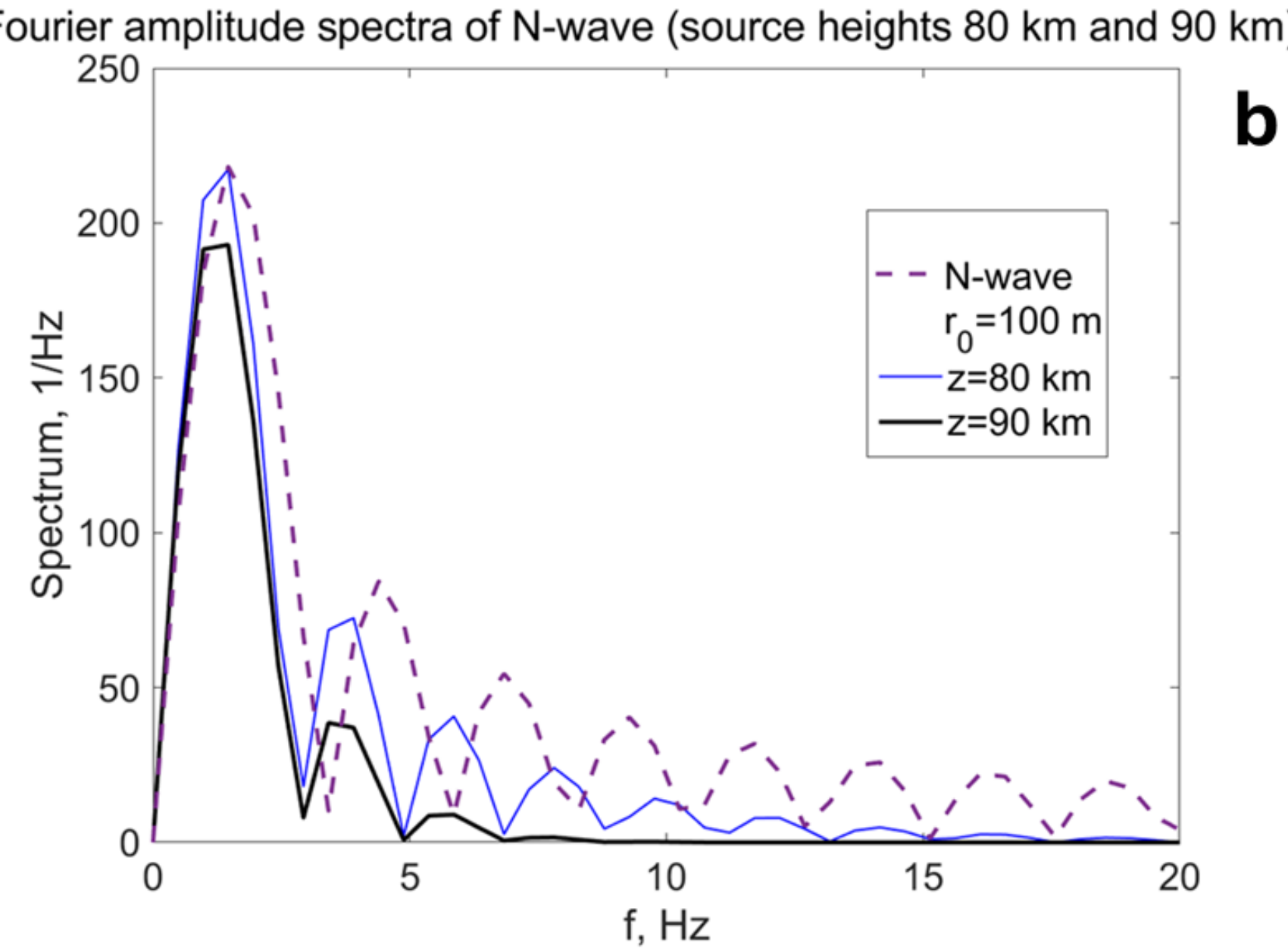

**Figure 8:** Transformation of the N wave with initial time duration $T_0$=0.5 s (dominant frequency $f_0$=2 Hz) at $r_0$ =100 m from a point source (Cases 1a-1c). (a) Dimensionless velocity $u(r,t) = \frac{v(r,t)}{\left(\frac{v_0\ r_0}{r}\right)}$ at $r$=5 km for source altitudes of 80, 90, and 100 km. The dotted line shows the initial N-wave profile defined for $-T_0/2$ < t < $T_0/2$. At 80 km ($Re$ ~ 55, Case 1a), the waveform is largely preserved with only modest nonlinear stretching and shock-front smoothing ($\tau_{shock}$ ~ 0.1 T_0). At 90 km (Case 1b), substantial smoothing is evident. At 100 km (Case 1c), the N-wave shape is lost entirely. (b) Fourier amplitude spectra at $r_0$ = 100 m (dashed) and r = 5 km (solid) for source altitudes of 80 and 90 km, showing progressive attenuation of high-frequency spectral components.

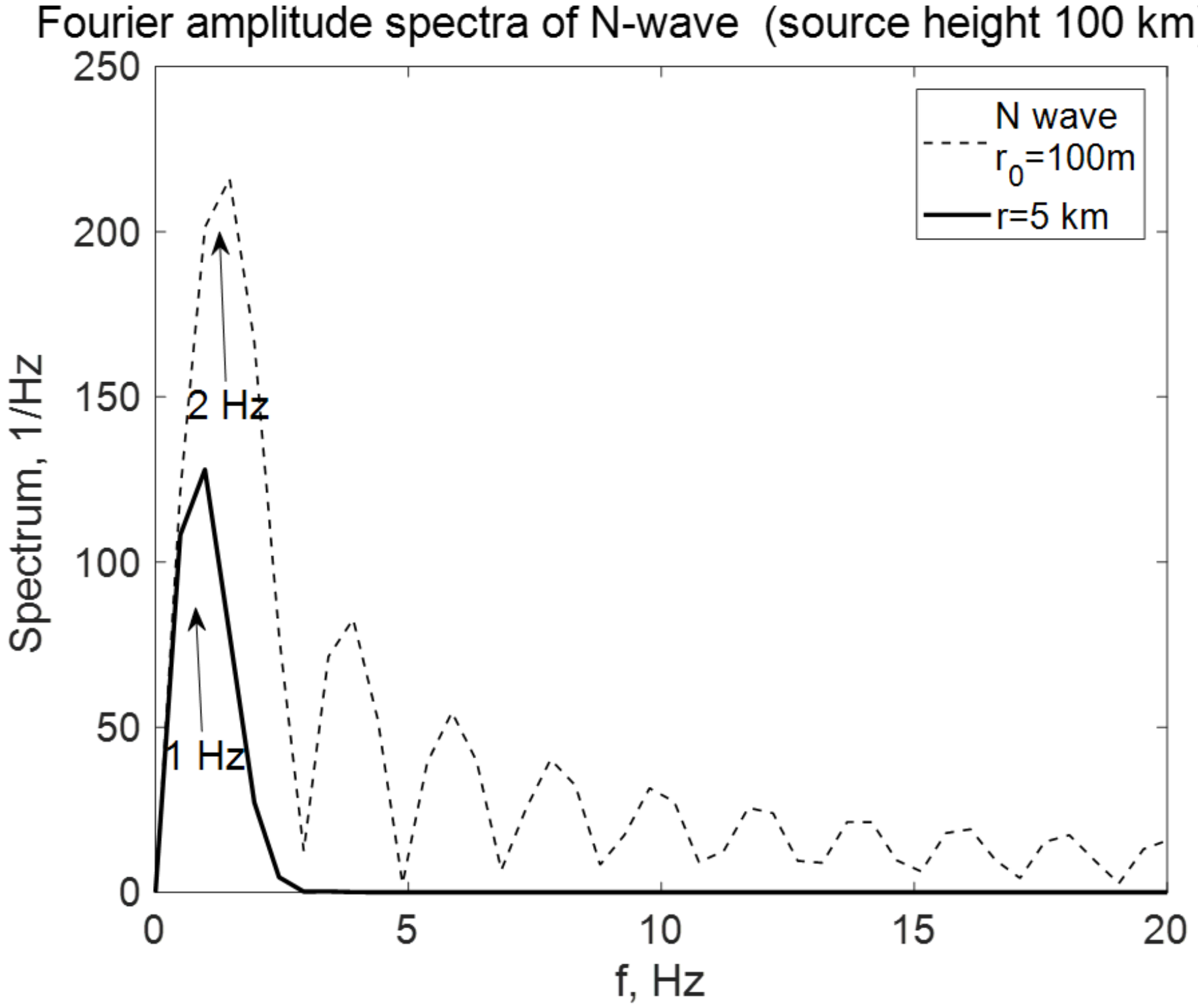

**Figure 9:** Fourier amplitude spectra of the N-wave signal ($T_0$ = 0.5 s, $f_0$ = 2 Hz) at $r_0$ =100 m (dashed) and $r$=5 km (solid) from their source at 100 km altitude (Case 1c). The strong attenuation of the high-frequency portion of the spectrum shifts the main spectral maximum from $f_0$ = 2 Hz down to approximately 1 Hz. Spectral components above 2 Hz are strongly absorbed, and the signal duration at $r$ = 5 km is approximately twice the initial N-wave duration.

#### 4.2.3 Case 2: Higher-Frequency Source, $f_0$ = 5 Hz at 90 km

For a shorter N-wave with $T_0$= 0.2 s ($f_0$ = 5 Hz) emitted at an altitude of 90 km (Case 2), strong absorption leads to significant smoothing of the signal profile and a shift of the dominant frequency from 5 Hz down to approximately 2.5 Hz at $r$ = 5 km (**Figure 10**). The waveform at r = 5 km is significantly smoothed and its duration becomes almost twice the initial dominant period of the N-wave. The surviving spectrum is concentrated at frequencies lower than 4 Hz. This supports the observational finding that receivers on the ground record dominant frequencies no higher than 3 Hz when meteoroids fragment at altitudes near 90 km (**Figure 6**).

At ranges $r$ > 5 km, further propagation of these signals with their surviving low-frequency spectra is affected mostly by refraction and scattering of spectral components by the fine-scale layered structure of the atmosphere crossing the ray paths. Parabolic equation calculations (Chunchuzov et al., 2025, Figures 2-4) show that this scattering does not significantly affect the modeled direct arrival on the ground up to horizontal ranges of approximately 200 km from the source. This is demonstrated by comparison of an initial signal at $r$ = 1.72 km from a source at $z$ = 89.6 km with its direct arrival on the ground at a horizontal distance of 172 km (Chunchuzov et al., 2025, Figure 5b,c): the low-frequency spectral structure is substantially preserved over the extended propagation path.

From the dissipation estimates for Cases 1b-1c, we can expect that for sources at heights near ~90 km, direct arrivals with dominant frequencies >2 Hz become rare at regional ranges. For a source at 100 km, direct arrivals are expected to be dominated by frequencies < 1 Hz at the ground. These predictions are directly confirmed by the frequency survival analysis (**Figure 6**).

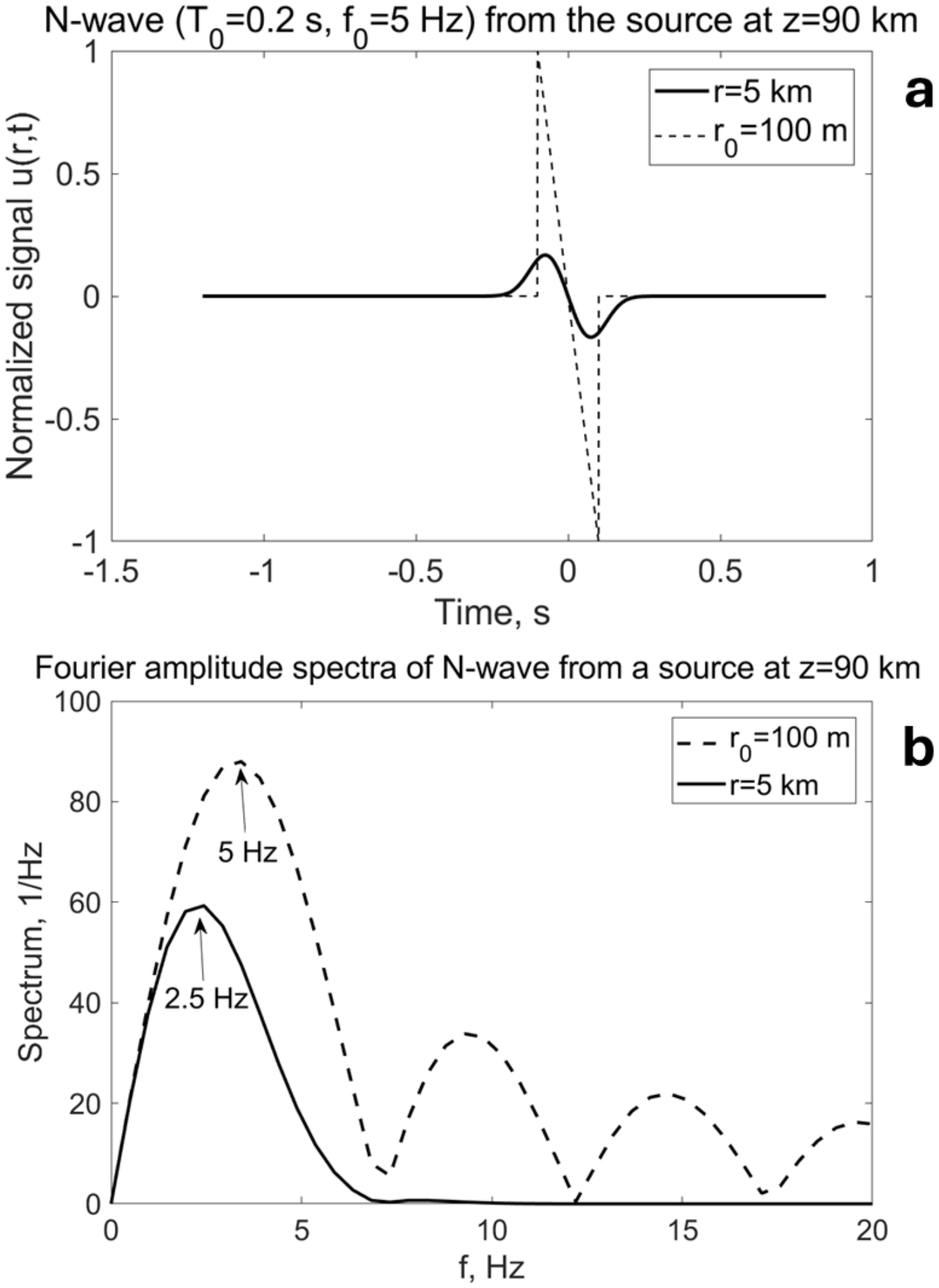

**Figure 10:** Transformation of a shorter N-wave with initial duration $T_0$= 0.2 s ($f_0$= 5 Hz) from a source at 90 km altitude (Case 2). (a) Waveform at $r$ = 5 km showing significant smoothing relative to the initial profile; the signal duration becomes almost twice the initial dominant period. (b) Fourier amplitude spectra at $r_0$=100 m (dashed) and $r$=5 km (solid). The dominant frequency shifts from 5 Hz to approximately 2.5 Hz, and the surviving spectrum is concentrated below 4 Hz.

**4.3 Regime Boundaries and Spectral Survival**

The observational trends and the near-source modeling converge on a consistent picture of altitude-dependent spectral filtering. The three $Re$ regimes introduced in **Section 2.3** provide the quantitative basis for linking the two:

i. Below ~80 km ($Re$ >> 1): nonlinear effects dominate, the N-wave shock front remains sharp (thickness ~0.1 $T_0$ from Eq. (8), and nonlinear duration stretching is modest (~1.1x). The emitted spectrum is largely preserved through the near-source region (Case 1a). In the dataset, 63% of detections below 80 km have dominant frequencies above 2 Hz and 31% above 3 Hz, consistent with the modeling prediction that spectral structure survives in this regime.

ii. 80-100 km (transitional): a transitional regime in which molecular dissipation begins to compete with nonlinearity. For a 2 Hz source, $Re$ = 1 is reached near 97-98 km. For a 5 Hz source, $Re$ = 1 is reached near 93-94 km. This means that the atmospheric filter preferentially removes higher-frequency content first, even at altitudes where the fundamental frequency is still in the weakly nonlinear regime.

iii. Above ~100 km ($Re$ < 1 for 2 Hz sources, $Re$ << 1 for higher frequencies): molecular dissipation dominates entirely, the problem reduces to linear frequency-dependent absorption, and the spectrum reaching the far field is confined to a progressively narrower low-frequency band (Case 1c). At 100 km, the modeling predicts that the surviving spectrum beyond 5 km is dominated by content below ~1 Hz; the observations are consistent with this, with 75% of detections above 100 km dominated by sub-1 Hz content and only a single detection above 90 km (out of 19) retaining a dominant frequency above 2 Hz.

The agreement between the modeled regime boundaries and the observed spectral survival thresholds is notably close. The modeling predicts attenuation of frequencies above ~4 Hz at 90 km and above ~2 Hz at 100 km; the data show $f$ > 3 Hz content vanishing above 80 km and $f$ > 2 Hz content becoming rare above 90 km. This close correspondence supports the interpretation that, for regional observations at the ranges considered here (< 300 km), near-source dissipation rather than path-integrated propagation effects is the primary driver of the observed altitude-frequency trends.

This interpretation is further supported by the partial correlation analysis. If the spectral trends were driven by propagation distance (longer paths providing more opportunity for frequency-dependent attenuation along the ray), the altitude-period correlation should weaken substantially after controlling for range. Instead, the partial correlation controlling for horizontal range ($r_s$ = +0.650) is virtually identical to the raw correlation ($r_s$ = +0.645), and controlling for three-dimensional range gives a similar result ($r_s$ = +0.558). This is what would be expected if the spectral modification occurs primarily in the near-source zone rather than along the extended propagation path.

The role of entry velocity in the observational SOMN–ELFO dataset warrants additional comment. Fast meteoroids (>40 km/s; $N$ = 40) generate sources at systematically higher altitudes (87.6 ± 9.6 km) than slow meteoroids (≤40 km/s; $N$ = 50; 69.9 ± 16.2 km), and the velocity-altitude correlation is strong ($r_s$ = +0.705). After controlling for velocity, the altitude-period partial correlation drops to $r_s$ = +0.360, still significant ($p$ = 5.0 x $10^{-4}$) but substantially reduced. This indicates that velocity contributes to the altitude–period trend observed in the SOMN–ELFO detections, primarily through its influence on source altitude, while the filtering mechanism retains independent explanatory power. In physical terms, faster meteoroids ablate and fragment higher, placing their acoustic sources in the dissipation-dominated regime, but the filtering itself is governed by the local atmospheric state (through $\nu$) rather than by any intrinsic property of the meteoroid.

### 4.4 Implications for Period-Based Yield Scaling

#### *4.4.1 Systematic Bias in Period-Based Energy Estimates*

The combined observational and modeling results carry direct consequences for period-based characterization of meteor infrasound. Period-yield relations (Eq. (1)) map a measured dominant period to a source energy through an empirically calibrated power law with exponents ranging from 3.30 to 3.75 across the principal published relations (Ens et al., 2012; Gi and Brown, 2017; ReVelle, 1997; Silber et al., 2025c). As discussed in **Section 2.1**, the steep power-law exponents mean that even modest changes in period map to large changes in inferred energy. Because the dominant period measured at the ground is systematically “lengthened” by the near-source filtering mechanism for high-altitude sources, the period-energy mapping becomes altitude-dependent in a way that is not captured by the standard relations.

The magnitude of the bias can be quantified directly from the observations. For a representative exponent of 3.5, a factor of 2 increase in observed period, comparable to the mean period ratio between the below-80-km and above-80-km populations (2.2x), corresponds to an apparent energy increase of $2^{3.5}$, approximately 11x. For sources above 100 km, where mean periods are roughly 4-5 times longer than those below 60 km, the implied bias reaches $4^{3.5}$, approximately 128x, to $5^{3.5}$, approximately 280x. The near-source dissipation modeling provides a physics-based basis for these numbers: a 2 Hz N-wave at 100 km is shifted to ~1 Hz purely through atmospheric filtering within 5 km, a factor of 2 in period corresponding to a ~11x energy bias if interpreted through a standard period-yield relation. At 90 km, a 5 Hz source is shifted to 2.5 Hz, again a factor of 2, again implying an order-of-magnitude bias.

While a portion of the observed period increase with altitude may reflect genuinely larger source energies at higher altitudes (through the velocity-energy relationship), the filtering mechanism alone is sufficient to account for the magnitude of the observed spectral shifts. The bias acts in one direction: the filter always lengthens the observed period, and period-yield relations always map longer periods to higher energies. The result is therefore a systematic positive bias in energy estimates inferred from uncorrected period-yield relations. Independent energy constraints would be required to determine the true event-by-event overestimate for the full dataset.

#### 4.4.2 Why Large Bolides Are Less Affected

The filtering mechanism described here is most consequential for small, regional meteoroids and less so for the large bolides against which period-yield relations were originally calibrated. As discussed in **Section 2.2**, large bolides deposit most of their energy at lower altitudes, typically in the 20-50 km range (Ronac Giannone and Silber, 2026), where the kinematic viscosity is orders of magnitude lower than in the upper mesosphere and $Re >> 1$. Their intrinsically low-frequency signals (dominant periods of several seconds, corresponding to sub-hertz frequencies) lie well below the cutoff frequencies imposed by the atmospheric filter at any altitude. A ~1 kt event with a dominant period of several seconds corresponds to frequencies that survive even the most severe dissipation at 100 km.

Regional meteors are the opposite case on both counts. Their sources frequently lie above 80 km (46% of the present sample, and 21% above 90 km), placing them in or beyond the transitional dissipation regime. Their intrinsic spectral content, being shifted toward higher frequencies due to their smaller source dimensions, falls within the frequency range most susceptible to the atmospheric filter. This combination of high-altitude sources with high-frequency content makes regional meteors uniquely vulnerable to the filtering effect and uniquely unsuited for uncorrected period-based energy inference.

The practical significance is amplified by the sheer abundance of these small events. Centimeter- to decimeter-scale meteoroids dominate the flux of extraterrestrial material entering Earth's atmosphere (Brown et al., 2002; Brown et al., 2013) and account for the overwhelming majority of detections at regional infrasound networks. If period-based methods produce systematically inflated energy estimates for these numerous events, the cascading effect on flux calculations, mass influx budgets, and statistical characterizations of the near-Earth object environment could be substantial.

The implications extend beyond meteoroids to other classes of high-altitude acoustic sources (**Section 2.4**). Reentering space debris, spent rocket bodies, and similar anthropogenic objects can produce infrasonic signals from altitudes comparable to or exceeding those of small meteors, particularly during early high-altitude breakup phases (e.g., Clemente et al., 2025; Hatty et al., 2026; Ishihara et al., 2012; Silber et al., 2024). Infrasound-based monitoring of such events for space situational awareness, reentry characterization, or treaty verification faces the same spectral filtering constraints and the same potential for period-based mischaracterization when source altitude is not accounted for. The regime boundaries and practical guidance developed here apply equally to any high-altitude infrasound source.

#### 4.4.3 Practical Guidance and Regime Boundaries

The results define a practical regime boundary for period-based interpretation, outlined below.

i. For source altitudes below approximately 75-80 km ($Re >> 1$), spectral content above 2-3 Hz survives the near-source environment, the receiver period retains a closer correspondence to intrinsic source properties, and period-based characterization is better supported (Case 1a). In this regime, existing period-yield relations may be applied with the usual caveats regarding calibration population and measurement methodology.

ii. For source altitudes between 80 and 100 km, the atmosphere progressively modifies the received spectrum, with higher-frequency content being removed at lower altitudes (Cases 1b, 1c and 2). Period-based estimates in this range should be interpreted with caution and, where possible, accompanied by independent constraints on source altitude or spectral quality flags. The presence of dominant frequencies above 2-3 Hz can serve as an empirical indicator that the signal has not been severely filtered.

iii. For source altitudes above ~100 km ($Re$ < 1), the near-source dissipation is severe and the received period is increasingly a property of the atmospheric filter rather than the source (Case 1c). Period-based energy estimates in this regime carry a systematic positive bias that can reach one to two orders of magnitude or more. In the absence of independent energy constraints, period-based values for events with source altitudes in this range should be regarded as upper bounds at best.

A complementary diagnostic is the spectral content itself. The complete absence of dominant frequencies above 3 Hz for any detection above 80 km, and above 2 Hz for all but one detection above 90 km, provides a practical quality criterion: if a regional meteor infrasound detection exhibits a dominant frequency below ~2 Hz and the source altitude is above 80 km, the period is likely to have been modulated by near-source dissipation and should not be used for uncritical energy inference. This guidance applies irrespective of the assumed shock production mechanism, since the mode of shock generation (ballistic cylindrical shock vs. fragmentation-driven spherical blast) is often not independently constrained for infrasound-only detections or for events with limited optical data.

**4.5 Limitations and Future Directions**

Several caveats apply to the present analysis. The near-source dissipation modeling assumes a point-source fragmentation geometry with a fixed initial Mach number ($M_0$ = 0.01 at $r_0$ = 100 m). While this is directly applicable to fragmentation-generated shocks in the dataset, events producing cylindrical line-source shocks will exhibit different amplitude decay and nonlinear waveform evolution. As discussed in **Section 3.3**, the frequency-dependent absorption that drives the spectral filtering is governed by the local atmospheric state rather than the shock geometry, thus the qualitative conclusions regarding the altitude dependence of the spectral filter are robust to this simplification. The observational trends further support this, as they emerge from the full dataset encompassing both source types. Nonetheless, future work incorporating cylindrical line-source models (e.g., ReVelle, 1976; Silber, 2026; Silber et al., 2015) would provide a more complete quantitative treatment of the near-source transformation for the full range of meteor shock geometries. The effective Mach number will also vary with event energy and fragmentation behavior. Because $Re$ scales linearly with $v_0$ (and hence with $M_0$), weaker events will enter the dissipation-dominated regime at even lower altitudes than the 80-90 km boundary identified here, while stronger events may preserve somewhat more spectral content to slightly higher altitudes. However, as noted in **Section 2.3**, the exponential increase in kinematic viscosity with height (approximately a factor of 10 per 10 km) ensures that the transition between regimes is always sharp, and the qualitative conclusions are insensitive to the precise choice of $M_0$.

The dataset, while well constrained and internally consistent, is drawn from a single geographic region (southern Ontario) and a single infrasound array (ELFO). The atmospheric structure, particularly the mesospheric temperature and wind profiles that influence viscosity and refraction, respectively, will differ at other latitudes and seasons. While the physics of viscous dissipation is universal, the precise altitude at which the $Re \sim 1$ transition occurs may shift by several kilometers depending on local atmospheric conditions. Furthermore, 90 detections, although sufficient to establish the statistical trends reported here, represent a modest sample size for robust calibration of altitude-dependent corrections. Larger datasets assembled from multiple regional networks across different geographic and seasonal conditions will be needed to refine the regime boundaries, quantify the scatter within each regime, and develop statistically robust altitude-corrected period-yield relations. Nevertheless, the consistency between the observational trends, the partial correlation analysis, and the near-source dissipation modeling strongly suggests that caution is warranted when applying period-based energy estimates to any regional meteor event with a source altitude above ~80 km.

The present analysis remains entirely in period/frequency space, deliberately avoiding the computation of source energies from the infrasound data. This choice is motivated by the circular reasoning that would arise from using period-based relations to compute energies while simultaneously arguing that those relations are unreliable for high-altitude sources. An independent energy approach, such as blast radius analysis from the weak-shock subset (Silber et al., 2015; Silber et al., 2026b) or photometric mass estimates, would enable direct quantification of the energy overestimation factor as a function of altitude. The 24-event weak-shock calibration subset within the SOMN-ELFO dataset provides a promising avenue for this comparison, but requires detailed modeling beyond the scope of the present work.

We also note that this work is intended as an exploratory, first-order assessment of altitude-dependent spectral filtering in a curated set of confirmed detections. We intentionally do not attempt to quantify how station/ambient noise (including wind noise and time-varying background), instrument/array processing choices, or signal strength/source amplitude (and associated SNR-dependent detectability) may influence which spectral components are detectable at the receiver or the recovered ‘dominant frequency.’ These effects may contribute to scatter and selection in the detected sample, and a dedicated treatment of station noise and detectability is beyond the scope of the present study.

The strong altitude dependence of the spectral filter suggests that future period-yield formulations could incorporate explicit altitude terms, for example through a relation of the form $\log_{10}(E) = A\log_{10}(\tau) + B + Cz$, where $z$ is the source altitude. Developing and validating such a formulation will require larger, consistently processed datasets that include both independent energy constraints and well-determined infrasound source heights across a broad altitude range. The present study provides the physical basis and empirical evidence for such an effort, by demonstrating that the atmospheric state at the source altitude constitutes a first-order control on the period measured at the ground.

An additional extension would be to pose the near-source filtering problem as a constrained inversion. Given independent source-altitude constraints, source-receiver geometry, atmospheric profiles, and a parameterized source model, the observed waveform or spectrum could in principle be used to estimate the pre-filtered source spectrum or an altitude-dependent correction to the measured dominant period.

Moreover, future work could extend this treatment by solving the generalized Burgers equation numerically along representative atmospheric profiles for both spherical and cylindrical spreading, while varying source altitude, initial frequency, and initial amplitude. Such calculations would provide quantitative maps of the expected ground-observed dominant frequency and would refine the regime boundaries and reference-distance choices identified in this first-order analysis.

**5. Conclusions**

This study analyzes 90 infrasound detections from 71 optically constrained regional meteoroid events recorded by the Southern Ontario Meteor Network and the Elginfield Infrasound Array to demonstrate that the upper atmosphere imposes a systematic, altitude-dependent spectral filter on infrasonic signals from high-altitude sources. The principal findings are:

i. Receiver dominant frequency decreases monotonically with increasing source altitude (Spearman $r_s$ = -0.629, $p$ = 3.3 x $10^{-11}$), with a corresponding increase in dominant period $T_{dsp}$ ($r_s$ = +0.645, $p$ = 6.7 x $10^{-12}$). This trend is robust to controls for propagation distance (partial $r_s$ = +0.650 after removing horizontal range dependence) and retains independent explanatory power after controlling for entry velocity (partial $r_s$ = +0.360), confirming that the spectral modulation is primarily an altitude effect driven by atmospheric conditions at or near the source.

ii. High-frequency spectral content is progressively and systematically extinguished with increasing source altitude. No detections above 80 km retain dominant frequencies above 3 Hz, only one detection above 90 km (out of 19) exceeds 2 Hz, and 75% of detections above 100 km are dominated by sub-1 Hz content. The dominant frequency distributions narrow markedly with altitude, collapsing from a broad, high-variance distribution below 60 km (mean 5.3 ± 3.7 Hz) to a tight, low-frequency band above 100 km (0.9 ± 0.5 Hz).

iii. Near-source dissipation modeling based on the generalized Burgers equation reproduces the sense and approximate magnitude of the observed trends. The acoustic Reynolds number, which governs the competition between nonlinear steepening and molecular dissipation, is both altitude- and frequency-dependent. The spectral pre-conditioning occurs within the first 5 km below the source and for regional observations at the ranges considered here (< 300 km), is consistent with being the primary driver of the ground-observed spectral trends.

iv. The modeled regime boundaries align closely with the observed spectral survival thresholds. The $f$ > 3 Hz survival fraction drops to zero at 80 km (consistent with the Re ~ 1 transitional regime onset), and the $f$ > 2 Hz fraction approaches zero above 90 km (consistent with the $Re$ << 1 linear dissipation regime). At 100 km, a 2 Hz N-wave is shifted to ~1 Hz within 5 km of the source (Case 1c), and a 5 Hz source at 90 km is reduced to ~2.5 Hz (Case 2). These modeled frequency shifts match the observed dominant frequency characteristics of the corresponding altitude populations in the dataset.

v. Because period-yield relations map $T_{dsp}$ to source energy through a power law with exponents of 3.3-3.75, the altitude-dependent filtering introduces a systematic positive bias in inferred energy for high-altitude sources. A factor of 2 increase in dominant period, comparable to the observed shift across the 80 km threshold, would imply an apparent energy overestimate of approximately one order of magnitude. For sources above 100 km, the corresponding apparent offset can reach two orders of magnitude or more. This effect is most consequential for small, centimeter-scale meteoroids, which constitute the most abundant class of atmospheric entry events and which routinely generate infrasound sources above 80 km.

vi. The results define three practical regimes for period-based inference: below ~80 km ($Re$ >> 1), where the emitted spectrum is largely preserved and period-based methods are better supported; 80-100 km, a transitional zone where caution and independent constraints are warranted; and above ~100 km ($Re$ < 1), where $T_{dsp}$ increasingly reflects the atmospheric filter rather than the source and period-based energy estimates should be regarded as upper bounds. The presence or absence of dominant frequencies above 2-3 Hz provides a complementary diagnostic for assessing the degree of near-source filtering. Because the mode of shock production (cylindrical ballistic shock vs. spherical fragmentation blast) is often not independently constrained, these regime boundaries represent an important practical consideration when interpreting dominant periods from any high-altitude source.

vii. The near-source dissipation mechanism identified here applies to any high-altitude acoustic source. Reentering space debris, spent rocket bodies, returning space missions, and other anthropogenic objects that generate infrasound from comparable altitudes are subject to the same physics. The regime boundaries and practical guidance developed in this work are therefore relevant to infrasound-based monitoring for space situational awareness and atmospheric entry characterization.

Our results demonstrate that the thermodynamic state of the atmosphere at the source altitude constitutes a first-order control on the dominant period of regional infrasound and must be accounted for in any period-based interpretation of high-altitude sources. Future work should pursue altitude-corrected period-yield formulations calibrated against independently constrained source energies, extend this analysis to additional regional networks to characterize seasonal and geographic variability of the dissipation regime boundaries, and incorporate cylindrical line-source modeling to provide a complete quantitative treatment across all meteor shock geometries.

**Data Availability**

The full SOMN–ELFO regional meteor catalog used in this study, including optical solutions, infrasound parameters, and derived quantities for the weak-shock calibration subset, is available via a Zenodo repository (https://doi.org/10.5281/zenodo.15868512).

**Acknowledgements**

This article has been authored by an employee of National Technology & Engineering Solutions of Sandia, LLC under Contract No. DE-NA0003525 with the U.S. Department of Energy (DOE). The employee owns all right, title and interest in and to the article and is solely responsible for its contents. The United States Government retains and the publisher, by accepting the article for publication, acknowledges that the United States Government retains a non-exclusive, paid-up, irrevocable, world-wide license to publish or reproduce the published form of this article or allow others to do so, for United States Government purposes. The DOE will provide public access to these results of federally sponsored research in accordance with the DOE Public Access Plan https://www.energy.gov/downloads/doe-public-access-plan. This paper describes objective technical results and analysis. Any subjective views or opinions that might be expressed in the paper do not necessarily represent the views of the U.S. Department of Energy or the United States Government.

**Funding**

EAS was supported by the Laboratory Directed Research and Development (LDRD) program at Sandia National Laboratories, project number 229346. IPC, OEP and SNK were partially supported by the Russian Science Foundation (RSF) grant № 25-17-00060 (Section 4.3).